\documentclass{aastex63}
\usepackage{multirow}
\received{}
\revised{}
\accepted{}
\submitjournal{ApJ}

\shorttitle{Evolution of the Orbital Light-curve of Her X-1 with 35-day Phase}
\shortauthors{Wang \& Leahy}
\graphicspath{{./}{figures/}}

\begin{document}

\title{The  Evolution of the Orbital Light-curve of Hercules X-1 with 35-day Phase}

\correspondingauthor{Denis Leahy}
\email{leahy@ucalgary.ca}

\author{Yuyang Wang}
\author[0000-0002-4814-958X]{Denis Leahy}
\affiliation{Dept. of Physics \& Astronomy, University of Calgary, University of Calgary, Calgary, Alberta, Canada T2N 1N4}



\begin{abstract}

Hercules X-1/HZ Hercules (Her X-1/HZ Her) is an X-ray binary monitored by multiple X-ray missions since last century. 
With the abundance of long-term observations, we present a complete set of 
orbital light-curves of Her X-1/HZ Her during the six states of the 35-day cycle in multiple energy bands. 
These illustrate in detail the changing light-curve caused by the rotating twisted-tilted accretion disc surrounding the neutron star.
The orbital light-curves during Main-High (MH) state are analyzed in 0.05 35-day phase intervals. 
These show the regular occurrence of pre-eclipse dips which march to earlier orbital phase as 35-day phase increase.
From the multi-band light-curves we derive time-average orbital phase dependence of column density for photoelectric absorption and energy-independent transmission as a function of 35-day phase. 
The X-ray light-curves during Low States are similar in shape to the optical Low State light-curve, but X-ray leads optical by $\simeq$0.04 to 0.08 in orbital phase. 

\end{abstract}



\section{Introduction}\label{sec:intro}

Her X-1/HZ Her is an X-ray binary of constant research interest for its abundant observational phenomena. 
The X-ray emission of the system features three timescales - the 1.24-second pulsation of Her X-1, the 1.7-day binary period, and the super-orbital 35-day cycle. 
The counter-precession of a twisted-tilted accretion disc is believed to be the reason causing the 35-day cycle \citep{1975ApJ...201L..61P, 1976ApJ...209..562G,1999ApJ...510..974S,2000ApJ...539..392S,2002MNRAS.334..847L}. 

Since its discovery in 1971 as an X-ray source, Her X-1 has been observed by various missions, accumulating large amount of data. 
The binary system has multi-wavelength emission including optical, ultraviolet (UV), extreme ultraviolet (EUV), and X-ray. 
UV and EUV radiation comes from the inner disc and the irradiated surface of HZ Her \citep{2000ApJ...542..446L,2020ApJ...889..131L} and is modeled by \citet{2003MNRAS.342..446L}. 
Optical emissions mainly originate from the X-ray heating of the companion star HZ Her by Her X-1. Modulation of optical light-curves of the 1.7-day orbits have been studied and give support to the precessing accretion disc model  \citep{1976ApJ...209..562G,1976ApJ...206..861D,2011MNRAS.418..437J}. 
In contrast, a recent study \citep{2020MNRAS.499.1747K} models the optical light-curves using a precessing neutron star and a forced disc.

Previous studies on the X-ray observations have shown properties of the 35-day cycle and pulse profile in detail. 
Long-term properties of the 35-day cycle of Her X-1 were studied by RXTE/ASM \citep{2011ApJ...736...74L} and Swift/BAT \citep{2020ApJ...902..146L} observations. 
The twisted-tilted counter-precessing disc model \citep{1975ApJ...201L..61P} has been developed to describe 35-day light-curve behaviours in multiple wavelengths \citep{2002MNRAS.334..847L}. 
The X-ray pulsations during Main-High (MH) were modeled with an accretion column by \citet{2004ApJ...613..517L}. 
\citet{2000ApJ...539..392S} modeled the evolution of the pulse profile of Her X-1 with 35-day cycle,  
 which was explained by the systematic changes in obscuration by the inner and outer disc edges. 
An alternative explanation for the change of the pulse shape with 35-day phase is  free precession of neutron star \citep{2013MNRAS.435.1147P}. 

Investigations on the timescale of 1.7-day binary period focus on the properties of eclipse, including ingress and egress, as well as dips, which are rapid drops in X-ray flux. 

Eclipse egress of Her X-1/HZ Her is rapid and less varying compared to the ingress.  
\citet{1988MmSAI..59..169O} show in their Fig. 7 the egress light-curves of several orbits observed by EXOSAT in 1984 and 1985. They found an increasing amount of absorbing material with 35-day phase, which is caused by the changing in X-ray illumination of the atmosphsere of HZ Her, or possibly by an extended structure of the outer edge of the accretion disc.   
\citet{1995ApJ...450..339L} studied five eclipses from GINGA/LAC observations between 1988 and 1989. Their spectra analysis indicates that during MH and early SH, most of the radiation from the neutron star is scattered without being absorbed, while in late high states, absorption dominates. 
\citet{1995MNRAS.276..607L} extracted the structure of atmosphere and wind around HZ Her by analyzing an X-ray ingress and eclipse during High State observed with the GINGA satellite in 1989. 
\citet{2014ApJ...793...79L} analyzed eight ingress and egress segments of high-resolution light-curves to measure the radius and evolutionary state of HZ Her. 
\citet{2015ApJ...800...32L} modeled the X-ray corona from electron scattering during High State X-ray eclipses. 

The existence of dips, and the fact that pre-eclipse dips march to earlier orbital phase were known since the very first observations of the system \citep{1973ApJ...184..227G, 1982ApJ...256..234G}. 
Previous studies of dips of Her X-1 include  \citet{1995A&A...297..747R} with EXOSAT data, \citet{2011MNRAS.418.2283I} with RXTE/PCA observations,
\citet{1999A&A...348..917S} and \citet{2012MNRAS.425....8I} in theory, and statistically - \citet{2011ApJ...736...74L}.

From the X-ray spectrum during dips, the dips are caused by absorbing material containing nearly neutral iron, thus nearly non-ionized (cold).  
Based on 44 sections of GINGA/LAC data in 1988 through 1990, \citet{1997MNRAS.287..622L} found the dip column density increases with orbital phase but not 35-day phase. They also found that there is no strong correlation between the duration of absorption dips and orbital or 35-day phase. 
Recent studies on Her X-1 made use of the RXTE/PCA data. 
The fraction of time in dip was found to vary with orbital phase, as well as 35-day phase \citep{2011ApJ...736...74L}.

However, the number of binary orbits investigated in these studies is relatively small. Often times only part of the orbital cycle is explored. 
On the other hand, X-ray monitors produce long-term observations of the system with significantly more orbital cycles but lower time resolution and sensitivity. 
As a result, monitoring data were analyzed with an emphasis on the super-orbital 35-day cycle. 
There exist no comprehensive analyses of the orbital light-curve of Her X-1/HZ Her in X-ray to date. 
In this paper, we present the time-average orbital light-curves derived from more than ten years of observations in multiple X-ray energy bands. 
We determine the change of orbital light-curves with 35-day phase. 
Further investigation of the pre-eclipse absorption dips is carried out during Main-High state.  

In Section \ref{sec:observations}, the X-ray observations are described. 
Section \ref{sec:analysis} describes the methods employed to create and fit the orbital light-curves. 
Section \ref{sec:results} presents results for the orbital light-curves during all six states of the 35-day cycle. 
Section \ref{sec:discussions} discusses the evolution of the orbital light-curve with 35-day phase and the physical implications. 
Section \ref{sec:conclusion} summarizes and concludes the paper. 

\section{Observations} \label{sec:observations}

\subsection{Swift/BAT} \label{subsec:BATobs}

The Burst Alert Telescope (BAT) on-board the Neil Gehrels Swift Observatory is an X-ray monitoring instrument, from which long-term light-curves are created \citep{2013ApJS..209...14K}. 
Data are available at the website \url{https://swift.gsfc.nasa.gov/results/transients/}, from which we downloaded the 15–50 keV light-curve of Her X-1. 
71165 points were included, covering 14 years of observation from February 2005 to November 2019. 
Exposure times per point range from 64 to 2672 seconds.
Time was recorded as mission time in unit of seconds, and was converted to MJD before further analysis. 

\subsection{RXTE/ASM} \label{subsec:ASMobs}

The Rossi X-Ray Timing Explorer (RXTE) All-Sky Monitor (ASM) monitored the sky in  the soft energy band of 2–12 keV with exposure time approximately 90 minutes \citep{1996ApJ...469L..33L}.
The Her X-1 light-curve is available at the Massachusetts Institute of Technology (MIT) RXTE project online database at \url{http://xte.mit.edu/asmlc/ASM.html}. 
The RXTE/ASM data cover the time period from 1996 January to 2011 December, with a total number of 95782 points.

\subsection{MAXI} \label{subsec:MAXIobs}

The Monitor of All-sky X-ray Image (MAXI) on the International Space Station monitors Her X-1 in the energy band of 2-20 keV \citep{2009PASJ...61..999M}. Simultaneous observations in three narrower bands are available for Her X-1: 2-4 keV (labeled as ``Band 1" in this paper), 4-10 keV (``Band 2"), and 10-20 keV (``Band 3"). 

For MAXI, data products (version 7L) were downloaded from the website provided by RIKEN, JAXA and the MAXI team
\footnote{\url{http://maxi.riken.jp/star\_data/J1657+353/J1657+353.html}}. 
A total number of 31927 observation intervals were obtained, covering the time period from 
August 2009 to June 2020. 
The exposure time of each interval is $\sim$90 minutes. 

\section{Analysis} \label{sec:analysis}

\subsection{35-day States} \label{subsec:P35d_calculation}

In order to study the relation between the orbital cycles of the binary system and the super-orbital 35-day cycle, 
we define the 35-day phase, in order that orbital light-curves during different 35-day states can be distinguished from each other. 

The time of X-ray Turn-On (TO) of 35-day cycles has been determined for more than 150 cycles in the last two decades \citep{2010ApJ...713..318L,2020ApJ...902..146L}. As a result, 35-day phase can be determined with high confidence level. 
We use the peak times of Swift/BAT 35-day cycles in Table 1 of \citet{2020ApJ...902..146L} to calculate the 35-day phase values for both Swift/BAT and MAXI observations. 

The definition of 35-day phase follows from \citet{2020ApJ...902..146L}, where the peak of Main-High state when flux reaches maximum is defined to be $\phi_{35d} = 0.0$ \footnote{Most earlier definitions of 35-day phase zero are ``start of TO", which is earlier by $\sim 0.13$ in 35-day phase than the definition used here.}.  
Under this definition, TO has 35-day phase $\phi_{35d, TO} = 0.87$ (Table 2 of  \cite{2021Univ....7..160L}). 
Here the data is split into the following ``states" of the 35-day cycle: 
mid Main-High (MH) $0.9 \le \phi_{35d} < 1.1$; 
decline of MH (DEC) $ 0.1 \le \phi_{35d} < 0.22$; 
first Low State (LS1) $ 0.22 \le \phi_{35d} < 0.4$; 
Short-High (SH) $ 0.4 \le \phi_{35d} < 0.65$; 
second Low State (LS2) $ 0.65 \le \phi_{35d} < 0.8$; 
and
rise of MH or turn-on (TO) $ 0.8 \le \phi_{35d} < 0.9$.

To better compare between the three missions, we analyze data with observation times which overlap with those of Swift/BAT, i.e. MJD53438-58692 (5255 days). 
As a result, RXTE/ASM data cover MJD53438-55923 of 2486 days, and MAXI data cover MJD55057-58692 of 3636 days.  
The total number of data points used for the study of orbital light-curves is 69806 for Swift/BAT, 34958 for RXTE/ASM, and 29061 for MAXI.

\subsection{Time-average Orbital Light-curves} \label{subsec:Porb_calculation}
We calculate orbital phase from observation time in MJD with Equation (5) of \citet{2009A&A...500..883S}. 
Mid-eclipse of the neutron star Her X-1 by its companion Her HZ defines orbital phase zero. 
For the different 35-day states, orbital light-curves were created for Swift/BAT, RXTE/ASM, and the three energy bands of MAXI. 

The data were binned in orbital phase.
The longest exposure time in Swift/BAT data is 2672 seconds (44.5 minutes), so we split the orbital light-curves of Swift/BAT into 50 bins such that every bin covers an orbital phase interval of 0.02, or $\sim49$ minutes. 
Because RXTE/ASM and MAXI have longer exposure times ($\sim90$ minutes), we reduce the resolution of the orbital light-curves to 25 bins. 
In each bin, if $N$ points are covered, the averaged count rate is the mean value $R_{bin} = \frac{1}{N} \sum_{i=0}^{N-1} R_i $, and the error is calculated as $\epsilon_{bin} = \frac{1}{N} \sqrt{\sum_{i=0}^{N-1} \epsilon_i^2} $.  

A total of 2946 orbital cycles are found in Swift/BAT data, 1286 in RXTE/ASM, and 1361 in MAXI. The average number of points averaged in each bin 
of the different 35-day states are summarized in Table \ref{tab:Npt_by_state}. 
The distribution of data between different states varies by a factor of $\sim2-3$.

\begin{table}[h]
\centering
\caption{Average Number of Points in Each Bin for Different 35-day States}
\label{tab:Npt_by_state}
\vspace{-2ex}
\begin{tabular}{c c c c c}
    \hline \hline
    State & $\phi_{35d}$ Range & Swift/BAT & RXTE/ASM & MAXI \\ 
    \hline
    MH & $0.9 \sim 1.1$ & 272 & 282 & 242 \\
    DEC & $0.1 \sim 0.22$ & 165 & 173 & 157 \\
    LS1 & $0.22 \sim 0.4$ & 257 & 260 & 207 \\
    SH & $0.4 \sim 0.65$ & 356 & 345 & 275 \\
    LS2 & $0.65 \sim 0.8$ & 209 & 206 & 174 \\
    TO & $0.8 \sim 0.9$ & 138 & 132 & 108 \\
    \hline
\multicolumn{5}{p{.45\textwidth}}{
\textbf{Note.} Ranges of 35-day phase are left-inclusive. 
}
\end{tabular}
\end{table}

\begin{figure*}
\gridline{\fig{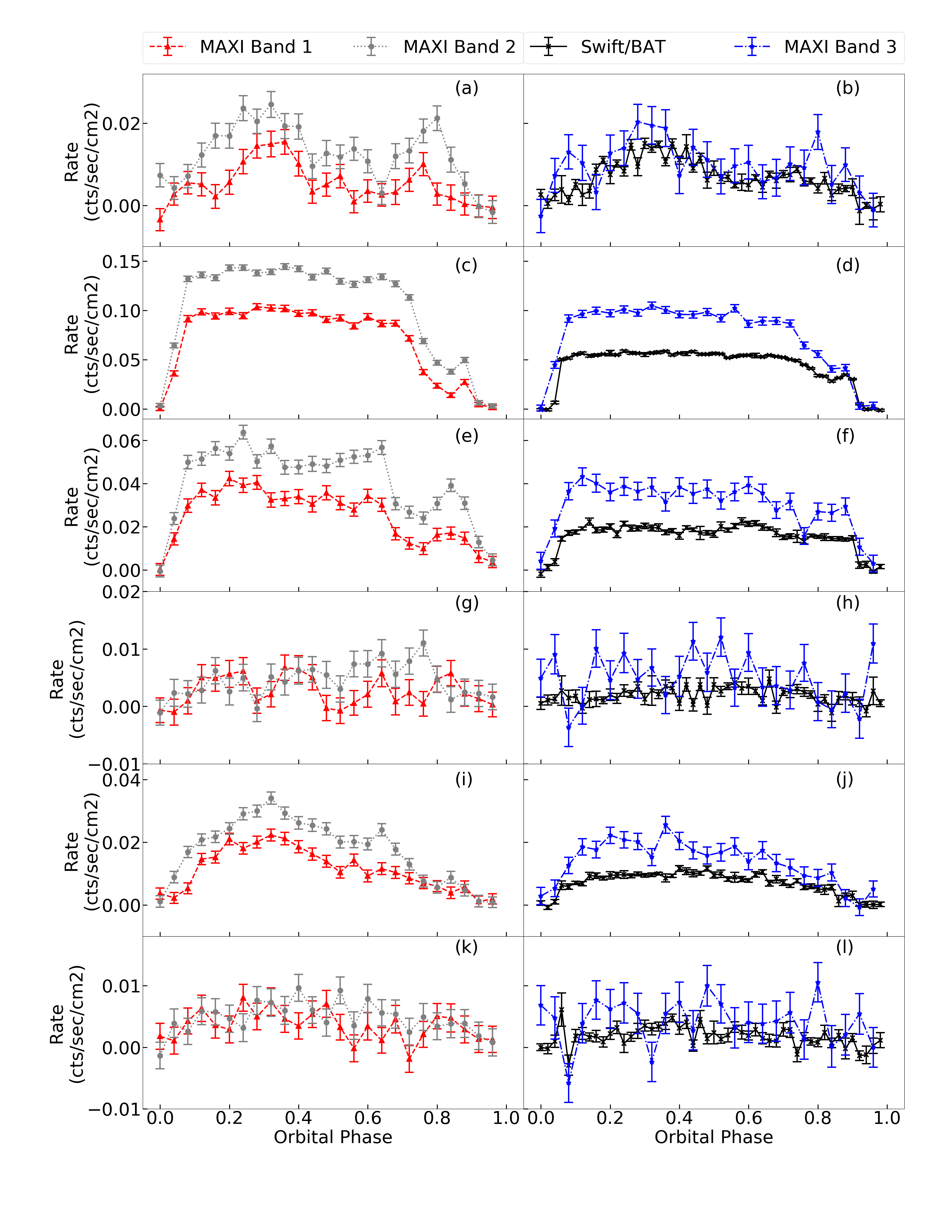}{\textwidth}{}}
\vspace{-15ex}
\caption{Orbital phase light-curves of Swift/BAT (50 bins) and three MAXI bands (25 bins) during all six 35-day states. (a)(b) TO ($ 0.8 \le \phi_{35d} < 0.9$); (c)(d) MH ($ 0.9 \le \phi_{35d} < 1.1$); (e)(f) DEC ($ 0.1 \le \phi_{35d} < 0.22$); (g)(h) LS1 ($ 0.22 \le \phi_{35d} < 0.4$); (i)(j) SH ($ 0.4 \le \phi_{35d} < 0.65$); (k)(l) LS2 ($ 0.65 \le \phi_{35d} < 0.8$). }
\label{fig:Orb_lc_BM123}
\end{figure*}

We show the orbital light-curves for the different 35-day states for Swift/BAT and 3 bands of MAXI in Figure \ref{fig:Orb_lc_BM123}. The 2-12 keV light-curves for RXTE/ASM are compared with those of MAXI in the Appendix (Figure \ref{fig:Orb_lc_AMtM12}).

\subsection{Light-curve Fits for Dips and Eclipses during MH and DEC}\label{subsec:fitMH}

During MH and DEC, dips and eclipse egresses and ingresses are clearly seen. 
The orbital light-curves during MH state show small enough errors that we
subdivide MH into smaller 35-day phase intervals. 
The MH is split into four equally-spaced 35-day ``sub-states": $0.9 \sim 0.95$, $0.95 \sim 1.0$, $0.0 \sim 0.05$, and $0.05 \sim 0.1$ (named MH-a, -b, -c, and -d, respectively). 
Table \ref{tab:num_orbit_substate} gives the number of orbits of data included in each of the sub-states.
We fit models to the light-curves for the sub-states of MH and for DEC to measure the dips, ingresses and egresses.

\begin{table}[h]
\centering 
\caption{Number of Orbits Averaged during MH Sub-states and Decline. }
\label{tab:num_orbit_substate}
\vspace{-2ex}
\begin{tabular}{c c c c c c }
    \hline \hline
    \# of orbits & MH-a & MH-b & MH-c & MH-d & DEC \\ 
    \hline
    RXTE/ASM & 64 & 64 & 63 & 62 & 163 \\
    MAXI & 65 & 68 & 77 & 79 & 185 \\
    Swift/BAT & 146 & 143 & 152 & 139 & 360\\
    \hline 
\multicolumn{6}{p{.45\textwidth}}{
\textbf{Note.} 
For RXTE/ASM and MAXI, 35-day phases are only calculated for the time periods overlapping with Swift/BAT, resulting in a smaller number of orbits. 
}
\end{tabular}
\end{table}

The eclipses during MH last from orbital phase 0.93 to 1.07 \citep{1995MNRAS.276..607L}. 
\citet{2011ApJ...736...74L} found the distribution of dips with respect to orbital phase (their Fig. 7). 
The fraction of time in dips is at minimum between orbital phase 0.2 and 0.4. 
Because the averaged light-curves during MH (see Section 4 below), show a  constant count rate in this range, we average over orbital phase 0.2 to 0.4
to determine a ``no-dip" count rate for each energy band, labelled $R_{max}$. 

\subsubsection{Eclipse Egress} \label{subsubsec:fit_2linear}

No dip is seen during orbital phase $0.0 \le \phi_{orb} < 0.4$,
so we can fit eclipse egresses independently of dips and ingresses.
During egress, the companion HZ Her moves away from
the line-of-sight to Her X-1 and results in an increase in count rate. 

We first fit the egress with a linear function (joined to 0 countrate during eclipse).
The linear function results in poor fits. 
We find that egress can be satisfactorily fit with two linear functions:
\begin{equation}
y_{eg} = 
   \begin{cases} 
      0 &  x < \phi_1 \\
      y_{eg1} = k_1 x + b_1  &  \phi_1 \le x < \phi_2 \\
      y_{eg2} = k_2 x + b_2  &  \phi_2 \le x < \phi_3 \\
      R_{max} &  x \ge \phi_3 \\
   \end{cases}
\end{equation}
The count rate starts to rise from zero at orbital phase $\phi_1$, which defines the start of the first linear function $y_{eg1}$.  
The two linear functions are joined at ($\phi_2$, $R_{ratio} R_{max}$).
The ''no-dip" rate $R_{max}$ is reached at orbital phase $\phi_3$ which defines the end of $y_{eg2}$. 
As a result, $k_1 = \frac{R_{ratio} \times R_{max}}{\phi_2 - \phi_1}$, $b_1 = - k_1 \phi_1 $, and $k_2 = \frac{(1 - R_{ratio}) \times R_{max}}{\phi_3 - \phi_2}$, $b_2 = R_{max} - k_2 \phi_3 $.

By inspection of the light-curves, we set $\phi_1 = 0.03$ for Swift/BAT which has 0.02 width bins, and $\phi_1 = 0.0$ for RXTE/ASM and MAXI which have 0.04 width bins. 
The three free parameters are fitted to the data: $\phi_2$ (orbital phase where the two linear functions join), $\phi_3$ (orbital phase where the count rate reaches $R_{max}$), and $R_{ratio}$ (ratio of count rate to $R_{max}$ at the join point). 

\subsubsection{Ingress and Pre-eclipse Dip} \label{subsubsec:fit_Gaussian}

Because the dips can occur during orbital phase $0.4 \le \phi_{orb} \le 1.0$,
part of the dips overlap ingresses. 
Thus we fit ingresses and dips at the same time.
We find satisfactory fits 
with a linear function for egress and Gaussian for dip. 
We make a further assumption of symmetrical eclipse so that $k$, the absolute value of slope, is the same as that for egress when fitting egress with a single linear function.  

The fit function for ingress is:  
\begin{equation}
y_{in} = 
   \begin{cases} 
      R_{max} & x \le \phi_0 - \frac{R_{max}}{k} \\
      y_{eg} = - k (x - \phi_0)  & \phi_0 - \frac{R_{max}}{k} < x < \phi_0 \\
   \end{cases}
\end{equation}
and the fit function for dip is:  
\begin{equation}
    y_{dip} = A_{dip} \exp{\bigg( \frac{(x - \mu)^2}{-2 \sigma^2} \bigg)} 
\end{equation}
Thus the fit function for orbital phase $0.4 \sim \phi_0$ includes both components: 
\begin{equation}
    y_{in,dip} = y_{in} \times (1 - y_{dip})
\end{equation}

The four free parameters are:  
$\phi_0 \in [0.91, 0.95]$; 
$A_{Dip} \in [0, 1]$; 
$\mu \in [0.7, 0.99]$; 
and
$\sigma \in [0.02, 0.12]$. 

\subsubsection{Additional Dip at Orbital Phase $\sim 0.6$ during MH-a} \label{subsubsec:fit_dip2}

An additional dip is visible near orbital phase 0.6, prior to the pre-eclipse dip, for all energy bands during the MH-a ($0.90 \le \phi_{35d} < 0.95$) sub-state. 
The same wide dip is visible during TO, but is not seen in later MH sub-states or DEC. 
For MH-a, we fit a second Gaussian function between orbital phase 0.4 and 0.8: 
\begin{equation}
    y_{dip2} = A_{dip2} \exp \bigg( \frac{(x - \mu_2) ^2}{-2 \sigma_2 ^2} \bigg)
\end{equation}
so that, for MH-a, the combined fit function is then: 
\begin{equation}
    y_{in,dip2} = y_{in,dip} \times (1 - y_{dip2})
\end{equation}
There are three additional parameters for the second dip:  
$A_{Dip2} \in [0, 1]$, 
$\mu_2 \in [0.4, 0.8]$, and 
$\sigma_2 \in [0, 0.4]$. 
We modify the lower limit of $\mu$ to be 0.85 and upper limit of $\sigma$ to be 0.05 in order to minimize the influence of the additional dip near orbital phase 0.6 on the fit of pre-eclipse dip. 

To find $1\sigma$ error of fit parameters, we calculated the $\chi^2$ values for various parameter sets. 
The upper and lower limits of any specific parameter $p$ are the maximum and minimum values of those with $\chi^2 \in [\chi^2_{min}, \; \chi^2_{min} + \Delta \chi^2]$ (with the method and $\Delta \chi^2$ specified in \cite{2002nrca.book.....P}). 

\subsection{Column Density and Transmission Fraction vs. Orbital Phase} \label{subsec:serial_model}

We quantitatively measure the effect of attenuating matter on the X-ray spectrum by fitting the count rates in different energy bands. 
This can be done for any orbital phase, thus determining the absorption vs. orbital phase, and how it changes with 35-day phase.
This can be done for the MH substates and DEC, which have small enough errors. 

The model assumption is that the observed count rate, compared to the un-attenuated rate vs. orbital phase, is a result of two factors.
One factor is photoelectric absorption and the second is energy-independent loss. 
The latter could be blockage by optically thick matter or scattering of flux out of the line-of-sight. 
The maximum rate is assumed to be the un-attenuated rate from the central source.
To reduce errors, the maximum $R_{max}$ is taken as the mean value of rates between orbital phases 0.2 and 0.4. 

As a result, the observed count rate $R$ can be expressed as: 
\begin{equation}
    \frac{R}{R_{max}} = f e^{-\sigma_{pe}(E) N_{H}} 
\end{equation}
where $R_{max}$ is the non-absorbed count rate, 
$N_{H}$ is the column density of the accretion stream, 
$\sigma_{pe}$ is the photo-electric absorption cross section,
and
$f$ is the transmission fraction caused by energy-independent loss. 

The photoelectric absorption cross-section is well known as a function of energy. 
We use WebPIMMS\footnote{https://heasarc.gsfc.nasa.gov/cgi-bin/Tools/w3pimms/w3\\pimms\_pro.pl} to find the averaged photoelectric cross-section for the X-ray energy band, including instrument response, of Swift/BAT and for MAXI Bands 1, 2 and 3. To do this calculation, we input the Her X-1 peak MH spectrum as the incident spectrum to convolve with the detectors' responses.  
The resulting effective cross-sections are:
MAXI Band 1 (2-4 keV) - $\sigma_{pe, M1} = 7.7 \times 10^{-24} \; \text{cm}^2$, Band 2 (4-10 keV) - $\sigma_{pe, M2} = 9.8 \times 10^{-25} \; \text{cm}^2$, Band 3 (10-20 keV) - $\sigma_{pe, M3} = 1.6 \times 10^{-25} \; \text{cm}^2$, and Swift/BAT (15-50 keV) - $\sigma_{pe, B} = 3.5 \times 10^{-26} \; \text{cm}^2$.

With these cross-sections, we solve for the column density $N_{H}$ and the transmission fraction $f$ at any given orbital phase with observed count rates from any two energy bands. 
Two independent calculations of $N_{H}$ and $f$ are carried out: one from MAXI Band 1 and Band 3, and the other from MAXI Band 1 and Swift/BAT.
We find consistent results, and that the latter gives smaller errors, so we show results using MAXI Band 1 and Swift/BAT below. 

\section{Results} \label{sec:results}

Figure \ref{fig:Orb_lc_BM123} shows the orbital light-curves for MH for
the six different 35-day states (TO, MH, DEC, LS1, SH and LS2) in the Swift/BAT and MAXI bands. 
For each 35-day state, the light-curve shapes are similar across the energy bands. 
By summing the count rates of MAXI Band 1 and 2, we obtain MAXI 2-10 keV light-curves to compare to those from RXTE/ASM (2-12 keV) (Figure \ref{fig:Orb_lc_AMtM12}).
The MAXI 2-10 keV and RXTE/ASM have consistent shapes: the normalization is different because of different sensitivities; and RXTE/ASM has larger background subtraction errors. 

\subsection{Low States Light-curves}

\begin{figure}
\gridline{\fig{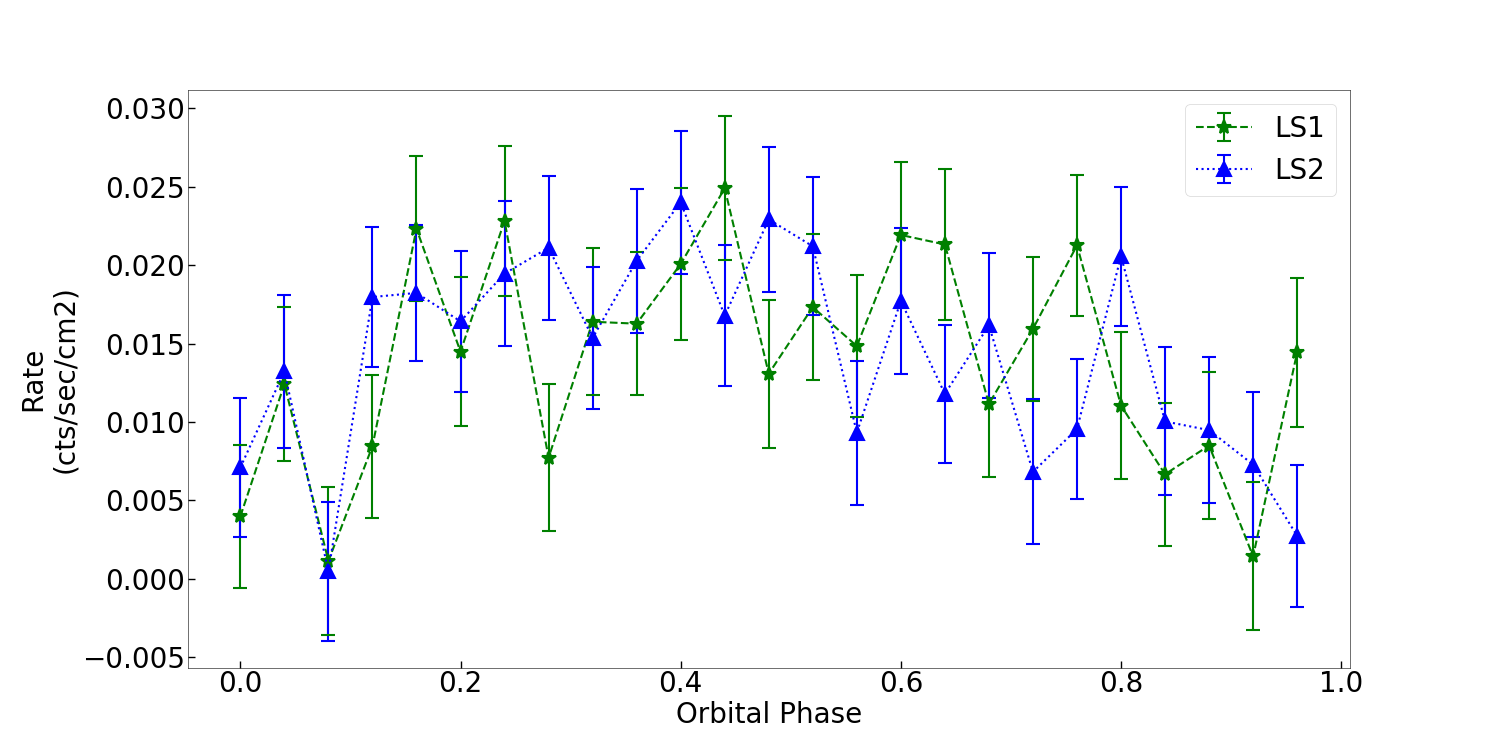}{0.5\textwidth}{}}
\vspace{-5ex}
\caption{Summed orbital light-curves during LS1 ($ 0.22 \le \phi_{35d} < 0.4$) and LS2 ($ 0.65 \le \phi_{35d} < 0.8$) from MAXI Band 1, 2, 3, and Swift/BAT. 
}
\label{fig:Orb_lc_LS_BM123}
\end{figure}

\begin{figure}
\gridline{\fig{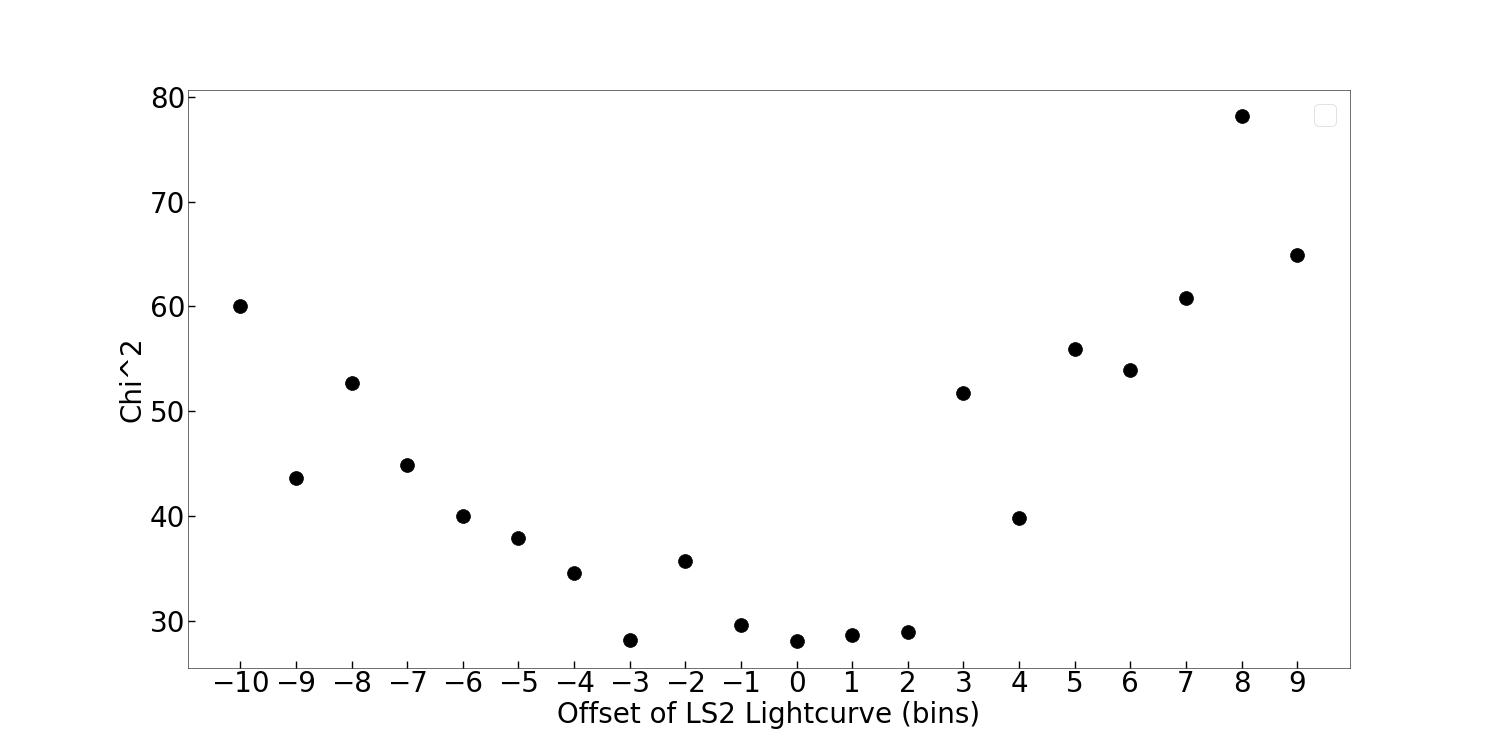}{0.5\textwidth}{}}
\vspace{-8ex}
\gridline{\fig{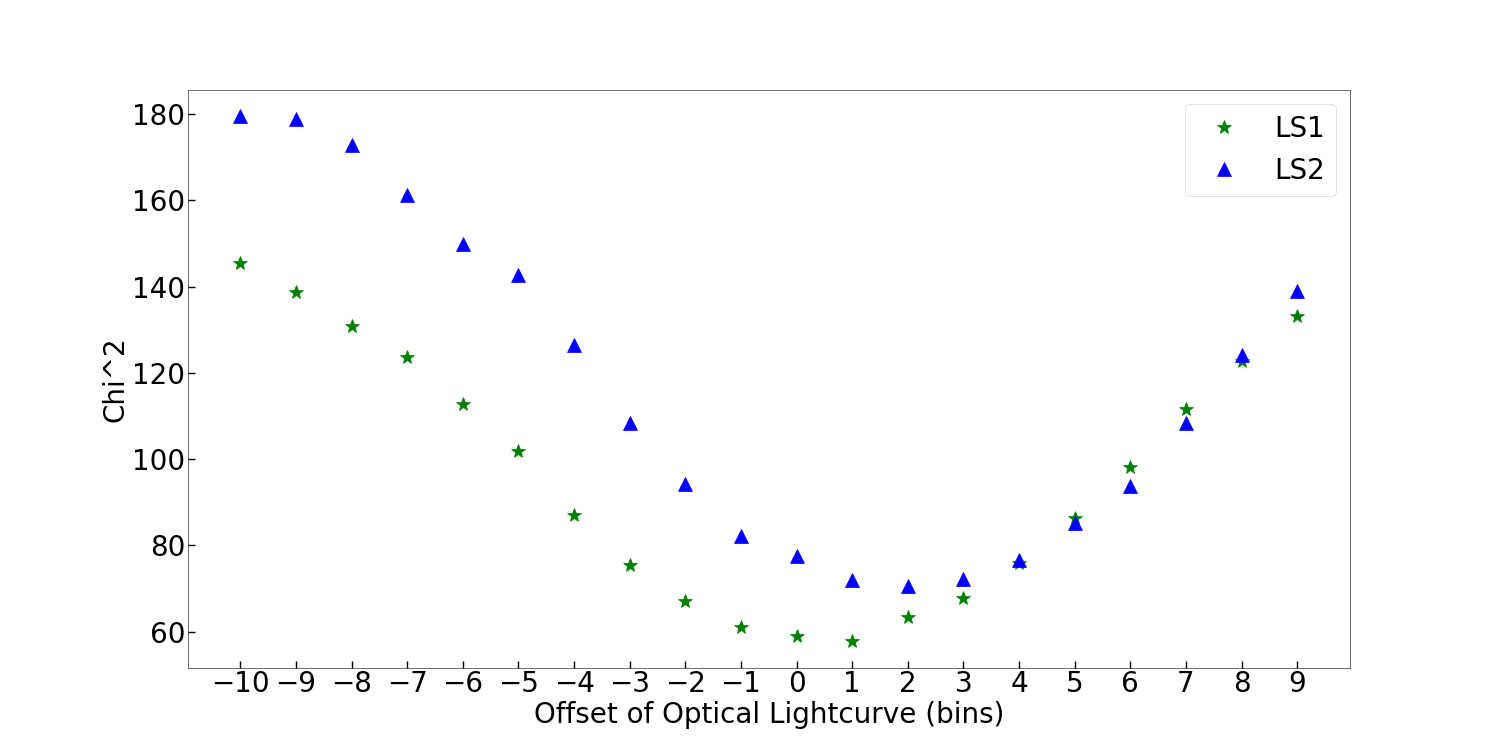}{0.5\textwidth}{}}
\vspace{-5ex}
\caption{$\chi^2$ vs. offset for LS1 and LS2.
Top panel: comparison of LS1 and LS2 orbital light-curves from Swift/BAT.
Negative number of bins means an offset of LS2 to later orbital phase than LS1.
Bottom panel: comparison of the LS1 and LS2 optical light-curves \citep{2011MNRAS.418..437J} with the sum of Swift/BAT and three MAXI bands light-curves of LS1 and LS2. 
The minimum $\chi^2$ occurs when the optical is shifted one bin earlier for LS1 (0.04 in orbital phase, or 1.63 hours) and two bins for LS2 (0.08 in orbital phase, or 3.26 hours). 
 }
\label{fig:Orb_lc_LSchi2}
\end{figure}

To reduce the X-ray data error bars for LS1 and LS2, we summed the Swift/BAT and 3 bands of MAXI light-curves (Figure \ref{fig:Orb_lc_LS_BM123}). 
The orbital light-curves during the two Low States exhibit little difference in shape, both showing a count rate of $\sim$ 4\% of the MH out-of-eclipse value (mean of $0.2 \le \phi_{orb} \le 0.6$). 
The relative count rate of LS with respect to MH peak value is consistent through all five energy bands, as well as  with other X-ray observations, e.g. 
BeppoSAX \citep{2000A&A...353..575O}, and  
AstroSat/SXT \citep{2019ApJ...871..152L}. 
A $\chi^2$ comparison test on the LS1 and LS2 light-curves from Swift/BAT 
yields a $\chi^2$ value of 28.1
which has probability level of 0.26. 
To test if there is an intrinsic phase offset between LS1 and LS2, 
we shift LS2 with respect to LS1, and calculate $\chi^2$. 
The minimum $\chi^2$ is at zero bins offset (Figure \ref{fig:Orb_lc_LSchi2}), showing no detectable offset.

A $\chi^2$ comparison test is performed on our X-ray orbital light-curves for LS1 and LS2 with the optical ones from \citet{2011MNRAS.418..437J}. 
To align their 35-day phase ranges with our two LS definitions, we average their light-curves over 35-day phase 0.35-0.50 for LS1, and 0.80-0.90 for LS2. 
The optical light-curves are binned to 25 bins to match the X-ray light-curve bins. 

The $\chi^2$ with no offset between X-ray and optical light-curves is 58.9 for  LS1, and 77.5 for LS2. 
However, we find smaller $\chi^2$ when a shift is included. 
The two light-curves are best aligned when optical is shifted 1 bin earlier than the X-ray (1.63 hours) for LS1, with $\chi^2_{min} = 57.8$, and 2 bins for LS2 (3.26 hours), with $\chi^2_{min} = 70.6$ (see Figure \ref{fig:Orb_lc_LSchi2} bottom panel). 

\subsection{Short-High Light-curve}

From previous X-ray observations, e.g. the 35-day light-curve from RXTE/PCA by \citet{2011ApJ...736...74L}, a typical peak rate of SH is about 25\% of MH. 
A similar value is seen in the 35-day light-curve derived from long-term analysis of Swift/BAT, e.g. bottom panel in Fig. 2 of \citet{2020ApJ...902..146L}.   
If we compare the SH peak count rate from our light-curves (in all bands) to that of the mean of MH and DEC, we find 29\%, 
which is comparable to previous studies. 

In all energy bands, the SH light-curve (Figure \ref{fig:Orb_lc_BM123} panels i, j) is asymmetric about orbital phase 0.5.
The decrease starts slightly before $\phi_{orb} \sim 0.4$, when HZ Her is behind Her X-1 and the accretion disc. 
In addition, this asymmetry is stronger in lower energy bands (MAXI Band 1 and 2), and weakest at the higher energy band of Swift/BAT which is least
affected by absorption. 
Thus, the decrease is likely due to absorption. This confirms results
from the study by \cite{2011ApJ...736...74L} which find
SH dominated by absorption dips after orbital phase $\sim0.4$.

\subsection{MH Light-curves} \label{subsec:fit_param}

The MH light-curves in different energy bands are shown in Figure \ref{fig:Orb_lc_BM123} (panels c and d).
The previous analysis to obtain a time-averaged MH orbital light-curve was done using
ASM data \citet{1999ApJ...510..974S} (their Fig. 4).
Our ASM MH light-curve is shown in Figure \ref{fig:Orb_lc_AMtM12} (panel b). 
It is consistent with the previous one, including the small rise at orbital phase 0.85 just before ingress. 
Ours has smaller errors: the previous one was constructed $\sim$850 days of data from MJD50146 to 50912, whereas ours in constructed from
$\sim$3250 days of data from MJD53438 to 55923.
 
Using the  four sub-states (MH-a, -b, -c, and -d) and DEC (MH decline) 
we examine the dependence of the light-curves on 35-day phase.

\subsubsection{Dip Parameters' Dependence on 35-day Phase}  \label{subsubsec:fit_result}

\begin{figure*}
\gridline{
        \hspace{-5ex}
        \fig{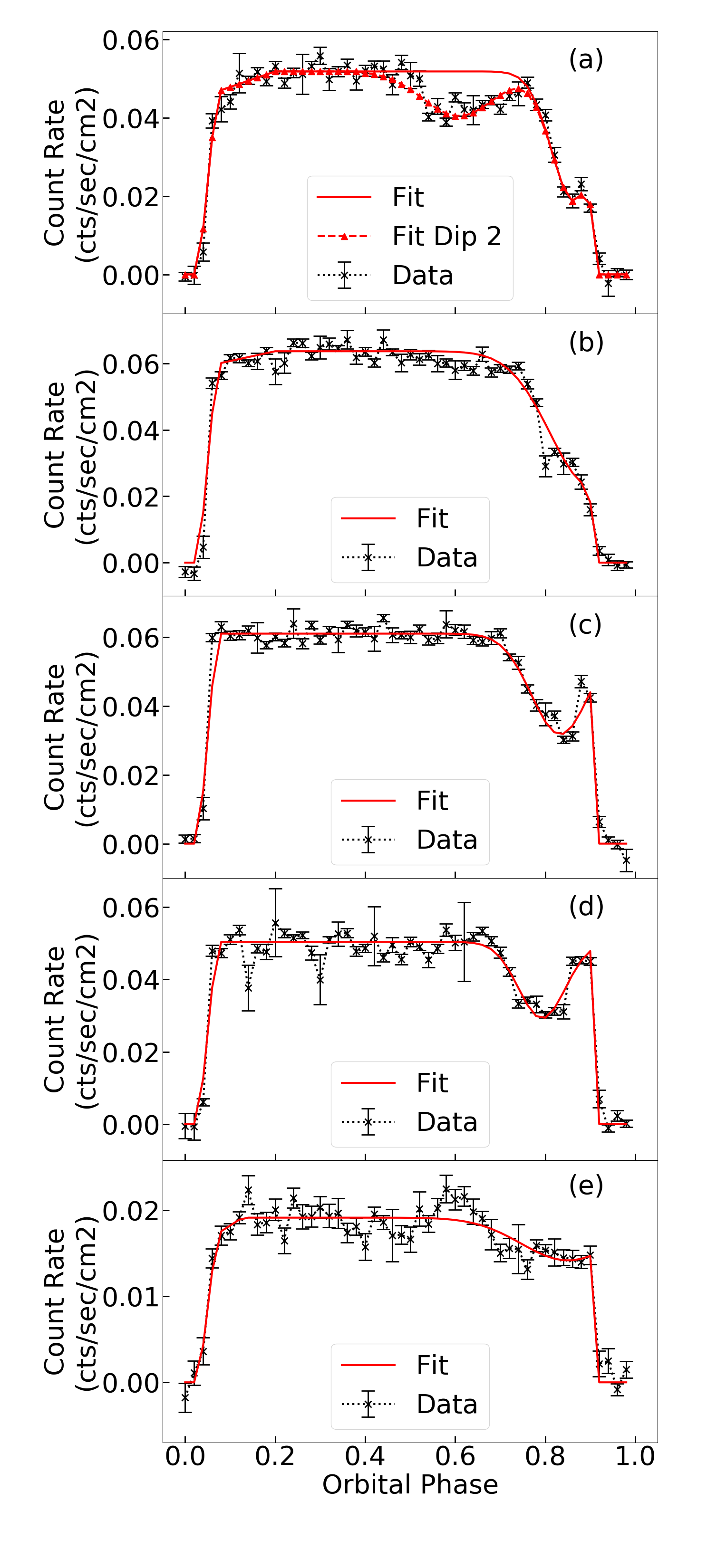}{0.53\textwidth}{}
        \hspace{-3ex}
        \fig{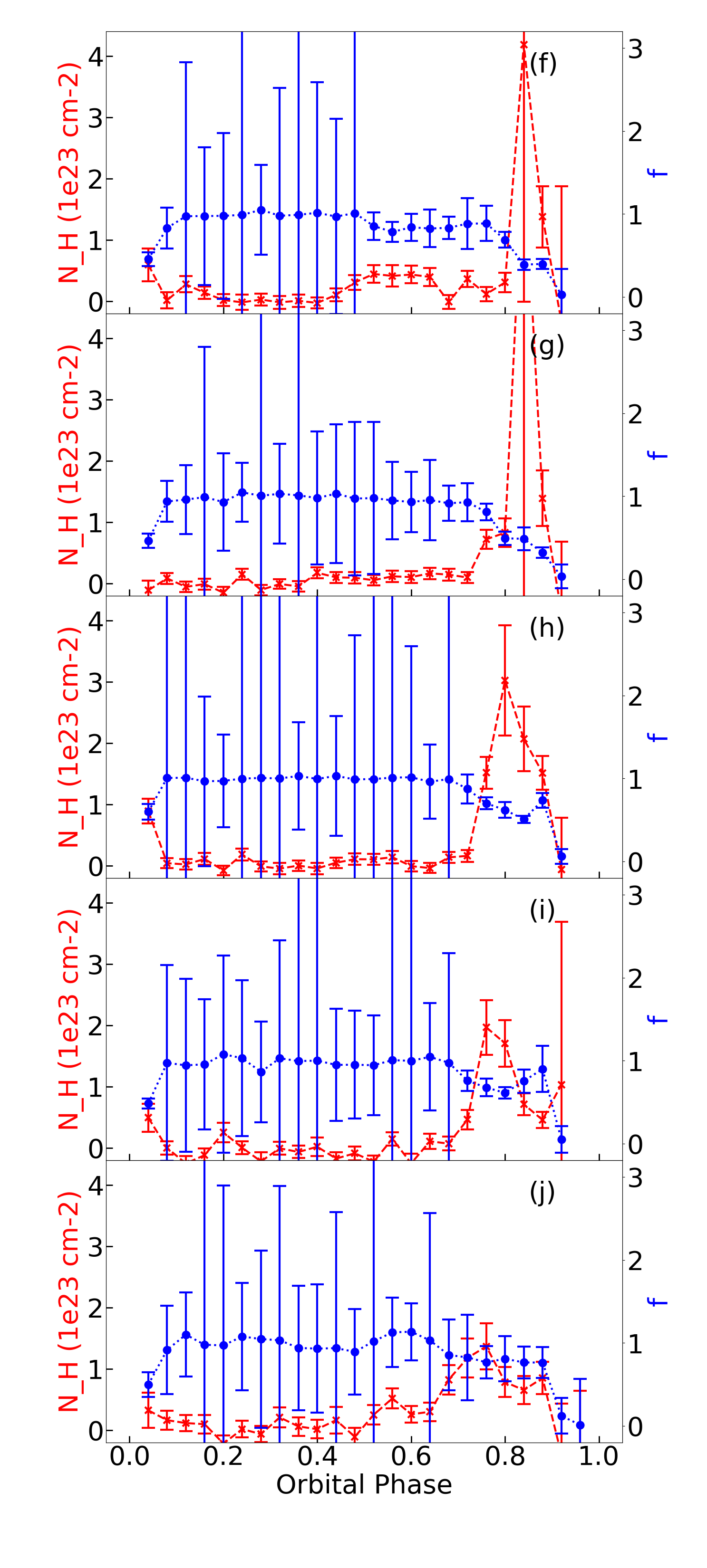}{0.53\textwidth}{}
          }
\vspace{-8ex}
\caption{Left column: best fit of orbital phase light-curves from Swift/BAT during MH and DEC (50 bins). 
See Appendix \ref{apdx:Orb_lc_fitMH_MA} for the plots of MAXI Bands 1, 2 and 3 and of RXTE/ASM.
Right column: column density $N_H$ (red crosses) and transmission fraction $f$ (blue dots) calculated from MAXI Band 1 (2-4 keV) \& Swift/BAT (15-50 keV). 
From top to bottom: 
(a)(f) MH-a ($ 0.9 \le \phi_{35d} < 0.95$); (b)(g) MH-b ($ 0.95 \le \phi_{35d} < 1.0$); (c)(h) MH-c ($ 0.0 \le \phi_{35d} < 0.05$); (d)(i) MH-d ($ 0.05 \le \phi_{35d} < 0.1$); (e)(j) DEC ($ 0.1 \le \phi_{35d} < 0.22$). }
\label{fig:Orb_lc_fitMH_B_NH_f}
\end{figure*}

The left column of Figure \ref{fig:Orb_lc_fitMH_B_NH_f} shows the orbital light-curves with the best fit egress-dip-ingress functions for the Swift/BAT data. The fits for RXTE/ASM and for the three MAXI bands give very similar 
results, and are included in Appendix \ref{apdx:Orb_lc_fitMH_MA}.  
There is an additional dip near orbital phase 0.6 preceding the pre-eclipse dip during MH-a. This dip will be discussed in Section \ref{subsec:dip_extra}. 

\begin{figure*}[h]
\gridline{\fig{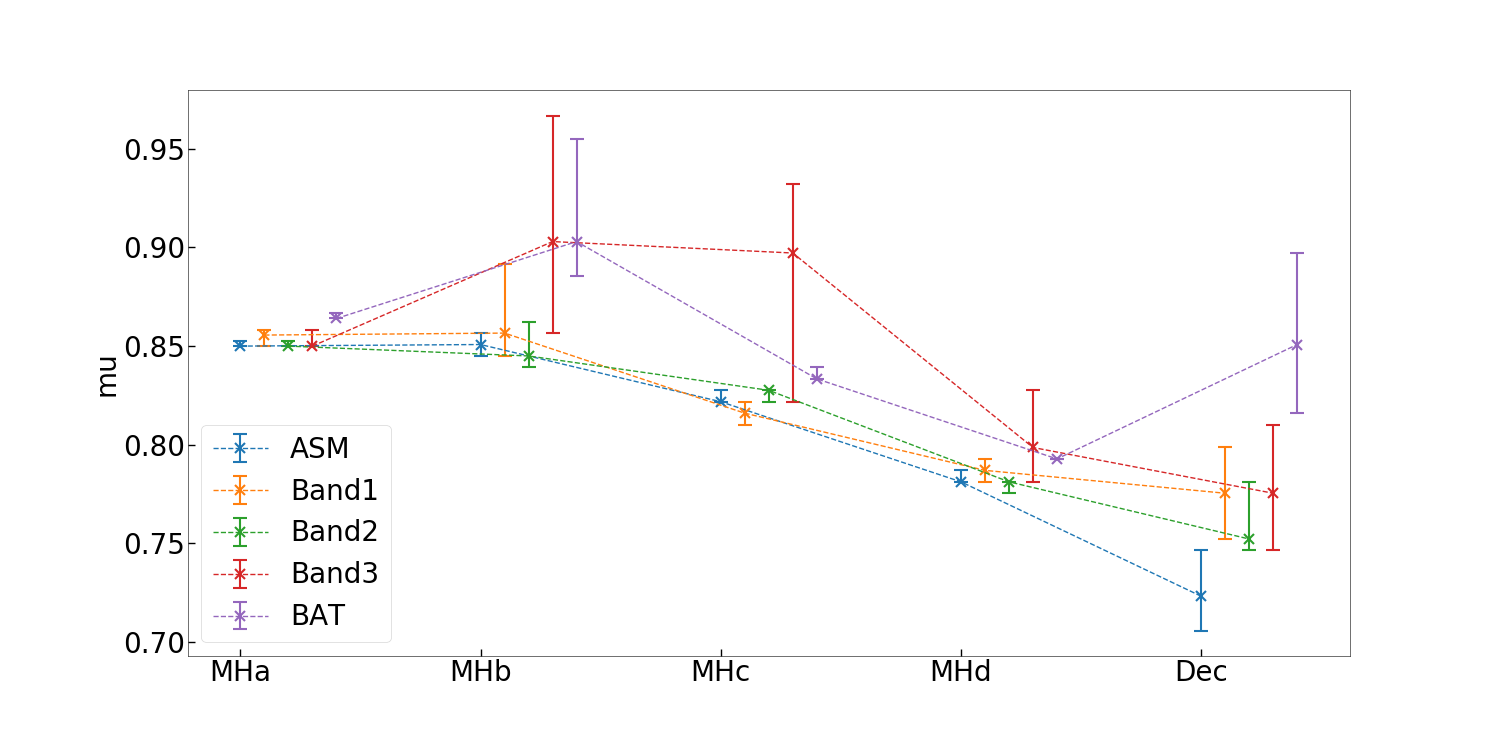}{0.55\textwidth}{(a)}
        \hspace{-7ex}
          \fig{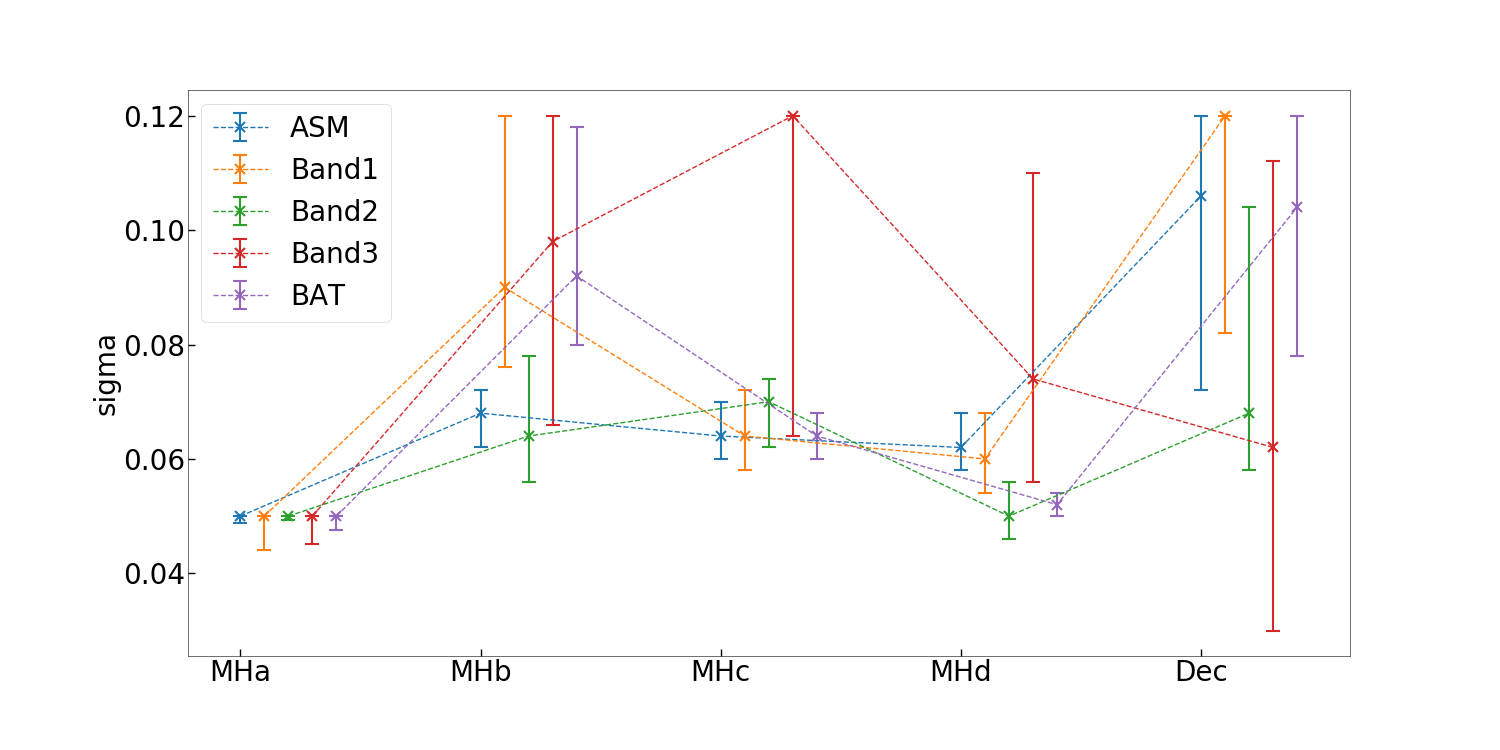}{0.55\textwidth}{(b)}
          }
        \vspace{-3ex}
\gridline{\fig{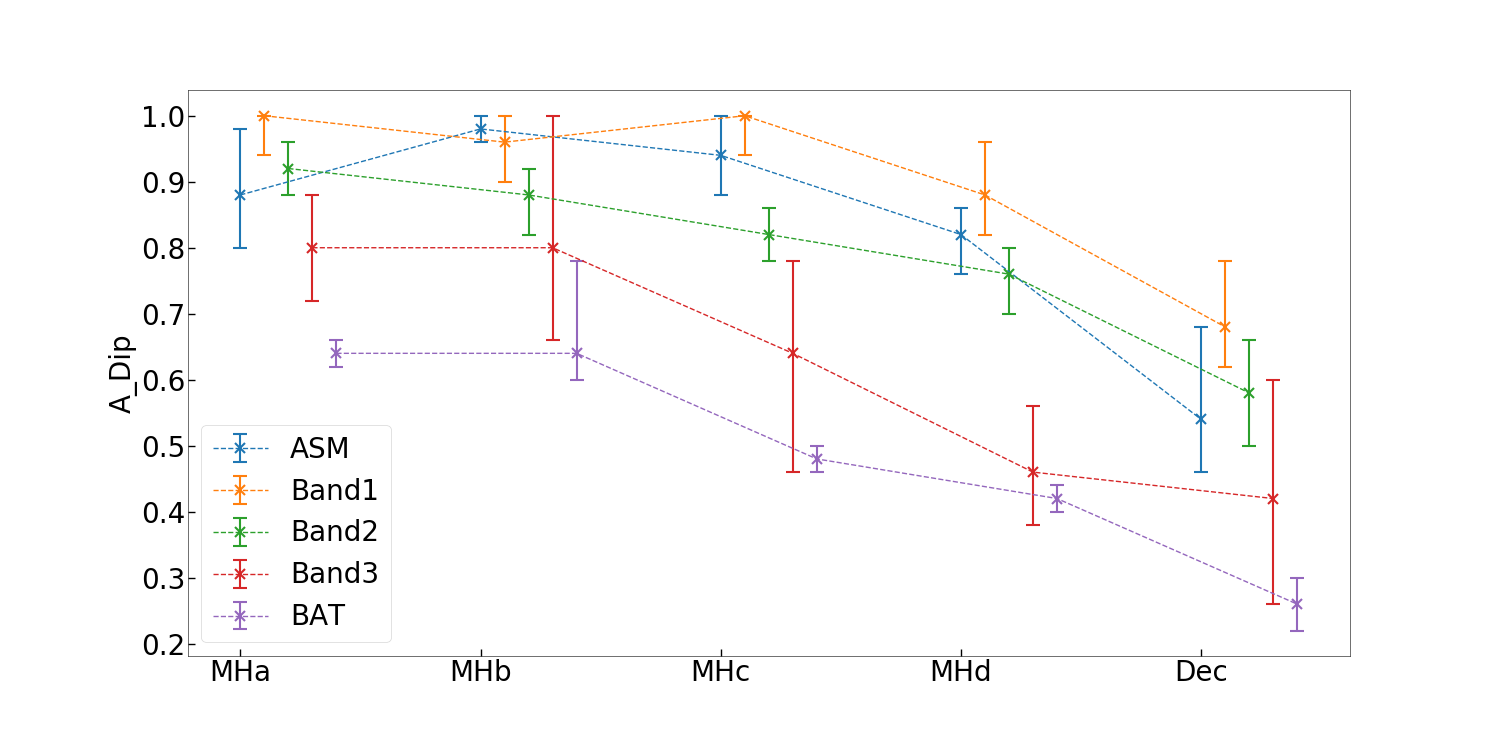}{0.55\textwidth}{(c)}
        \hspace{-7ex}
          \fig{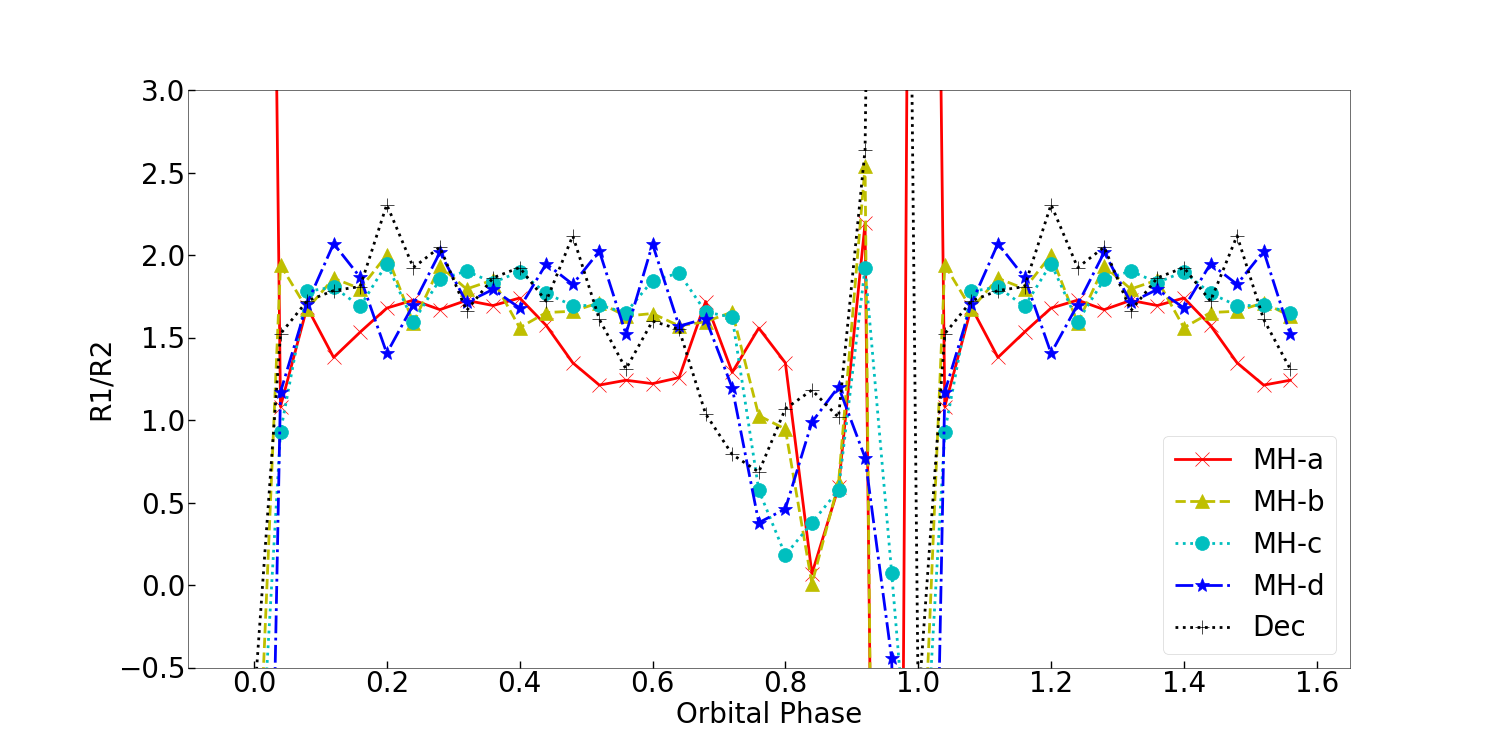}{0.55\textwidth}{(d)}
          }
\caption{Fit parameters of orbital light-curves during MH. 
(a) Centre of dip ($\mu$); 
(b) Width of dip ($\sigma$); 
(c) Depth of dip ($A_{dip}$); 
(d) Softness ratio ($R_1 / R_2$) of 1.5 orbits for MH sub-states and DEC, where $R_1=$ MAXI Band 1 (2-4 keV) and $R_2=$ Swift/BAT (15-50 keV). 
The errors of SR are $\simeq \pm 0.2$, except during eclipse for which they increase to $\sim 2$ and higher. }
\label{fig:Orb_lc_fit_param}
\end{figure*}

The parameters of best fit are listed as tables in Appendix \ref{apdx:fitMH_param}, along with the $\chi^2$ values. 
Errors are calculated as described in Section \ref{subsubsec:fit_Gaussian}. 
In Figure \ref{fig:Orb_lc_fit_param} we plot $\mu$, $\sigma$, and $A_{dip}$ vs. orbital phase. 
In addition, the softness ratio (SR) between  MAXI Band 1 (2-4 keV; $R_1$) and Swift/BAT (15-50 keV; $R_2$) is plotted in panel d of Figure \ref{fig:Orb_lc_fit_param} for the MH sub-states and for DEC. 
The drop in SR indicates photoelectric absorption, where soft X-rays are more strongly reduced than hard X-rays. 

The best fit functions for the MH sub-state light-curves are shown as the solid lines in the left column of Figure \ref{fig:Orb_lc_fitMH_B_NH_f}) with best fit parameters and their errors shown in panels a, b and c of Figure \ref{fig:Orb_lc_fit_param}).
For all energy bands, the pre-eclipse dips phase shifts to to earlier orbital phase ($\mu$ decreases, panel a) and become shallower ($A_{dip}$ decreases, panel c). 

As energy increases (Band 1 to Band 2 to Band 3 to BAT), there is marginal evidence that the dips occur later in orbital phase (panel a). 
There is stronger evidence that the dip depth decreases with energy (panel c),
as would be expected if the dips are mainly caused by absorption.
No significant evidence is found that the width of the dip depends on 35-day phase or energy (panel b). 
The mean value of dip width is $\sigma_{mean} = 0.07208$, or 2.9 hours, and the standard deviation is 0.02203 (0.90 hours). This gives the FWHM of the pre-eclipse dip in orbital phase: $2 \sqrt{2 \ln{2}} \; \sigma_{mean} = 0.1697$
(6.9 hours), comparable to the duration of the eclipse.

\subsubsection{Parameters for the Dip at Orbital Phase 0.6 during MH-a} \label{subsec:dip_extra}

During MH-a state, there is an additional absorption dip around orbital phase 0.6. 
The Gaussian fit shows that the depth decreases as energy increases, indicating absorption process, 
while the position and width of the dip in orbital phase are consistent among different energy bands. 

This wide absorption dip during early MH is much shallower than the pre-eclipse dip, and is not symmetrical in orbital phase. 
As an alternate approach, we approximate its depth by manually selecting a middle point where the count rate is $R_{dip}$ and compare to $R_{max}$. 
The percentage depth of the dip is then defined as: $ \% Depth = 1 - \frac{ R_{dip} }{ R_{max} }$. 
Table \ref{tab:dip_extra_depth} lists the middle point chosen and the corresponding percentage depth for all five energy bands, which is consistent with the results from the best Gaussian fit. 
MAXI and RXTE/ASM agree with each other with a depth of 30 - 40 \% of the averaged maximum count rate. Swift/BAT gives much lower value of 18 \%.
This means there is combined absorption and energy-independent blockage during the dip. 

\begin{table}[h]
\centering 
\caption{Approximated Percentage Depth and the best Gaussian fit parameters of the Absorption Dip at $\phi_{orb} \sim 0.6$ }
\label{tab:dip_extra_depth}
\vspace{-1ex}
\begin{tabular}{p{0.12\textwidth} p{0.04\textwidth} p{0.07\textwidth} p{0.04\textwidth} p{0.04\textwidth} p{0.04\textwidth}}
    \hline \hline
     & $\phi_{dip}$ & \% Depth & $A_{dip2}$ & $\mu_2$ & $\sigma_2$ \\ 
    [-2ex] \hline 
    RXTE/ASM & 0.56 & 39.53 \% & 0.45 & 0.60 & 0.08\\
    MAXI Band 1 & 0.56 & 39.45 \% & 0.40 & 0.60 & 0.10 \\
    MAXI Band 2 & 0.56 & 35.42 \% & 0.30 & 0.58 & 0.10 \\
    MAXI Band 3 & 0.64 & 31.94 \%  & 0.30 & 0.62 & 0.08\\
    Swift/BAT & 0.62 & 17.58 \% & 0.20 & 0.60 & 0.08\\
    \hline
\end{tabular}
\end{table}

The orbital phase when this wide dip appears does not show strong dependence on energy in our analysis because of two major limitations in data. 
Although long-term observations provide full coverage of data in orbital light-curves, the variation among individual cycles is lost in the process, and longer exposure time restricts the resolution in orbital phase. 
In addition, large errors (see Figure \ref{fig:Orb_lc_BM123}) during 35-day states other than MH makes it not possible to identify the wide dip during TO and even earlier 35-day phases. 

\subsection{Column Density and Transmission Fraction vs. Orbital Phase}

We show results for $N_{H}$ and $f$ determined from MAXI Band 1 and Swift/BAT in Figure \ref{fig:Orb_lc_fitMH_B_NH_f} (right column). 
Despite the large errors, the transmission fraction is consistent with unity for orbital phases between 0.1 and 0.6 to 0.8, depending on 35-day phase. 
The column density is small except during dips. 
During the pre-eclipse dip, we see a clear rise in the column density to a few times of $10^{23} \; \text{cm}^{-2}$, and a drop in the transmission fraction $f$. 
This can be explained by an increased absorption in the system that is accompanied by energy-independent losses. 

The out-of-dip column density, except near eclipse where the count rates are too low to measure $N_{H}$ and $f$, is $\sim10^{21} \; \text{cm}^{-2}$. 
During MH-a, the maximum column density of the extra absorption dip is $4.5 \times 10^{22} \; \text{cm}^{-2}$ at orbital phase 0.52. 
After MH peak, the maximum column density in pre-eclipse dip drops from $3.0 \times 10^{23} \; \text{cm}^{-2}$ at orbital phase 0.8 during MH-c to $1.4 \times 10^{23} \; \text{cm}^{-2}$ at orbital phase 0.76 in DEC. 

\section{Discussion} \label{sec:discussions}
\subsection{Low States}

We find an orbital phase delay of the optical light-curves during LS1 and LS2 compared to the X-ray light-curves of LS1 and LS2.
An X-ray - optical offset has not been reported before, and is possibly due to the different disc shadow affecting the illumination region on HZ Her. 

Despite the limitations of different methods, there is consistency in shape between our orbital light-curve and those measured in previous X-ray studies. 
Thus the LS orbital light-curves are stable over time periods of $\sim$30 years.

Limited by the sensitivity resolution of data, we don't clearly see eclipse in our LS light-curves, unlike some other studies. 
E.g., \citet{2015MNRAS.453.4222A} used RXTE/PCA data clearly see eclipse and found that the LS X-ray emission is mainly from reflection off the face of HZ Her and from the accretion disc corona (their Fig. 5). 
More recently, \citet{2021A&A...648A..39S} studied two observation periods during LS of Her X-1/HZ Her from the SRG/eROSITA all-sky survey (0.2-8 keV). 
They proposed a similar model of LS X-ray emission, but further break the reflection component from the companion star into an optically thick cold atmosphere and an optically thin hot corona above. 

\subsection{Short-High State}

In SH, the observer views the inner rings of the disc, and the system emits soft X-rays ($\sim1$ keV) mainly from the inner disc  \citep{1982ApJ...262..301M, 2002MNRAS.334..847L}. 
Harder X-rays ($>$1 keV, which we measure) are mainly from scattering by the inner disc and material above the disc, while the central neutron star source of the X-rays is obscured by the inner disc edge  \citep{2000ApJ...539..392S, 2002MNRAS.334..847L}. 

The decrease of X-ray emission for SH starts at orbital phase 0.35 (Figure \ref{fig:Orb_lc_BM123}), thus it cannot be caused by the accretion disc which rotates with 35-day period.
The companion is not in the observer's line-of-sight over the full orbital phase interval of 0.07 to 0.93. 
However, the accretion stream leads the companion in its orbit (see Fig. 1 of
\cite{2012MNRAS.425....8I}) and the accretion stream-disc interaction site
was proposed by that work as the cause of dips, including the 35-day and orbital phase dependence of the dips. 

A detailed picture of the orbital light-curve during SH was given by 
\citet{2015MNRAS.453.4222A}, with four successive orbits of high-sensitivity RXTE/PCA data (their Fig. 1). 
The first orbit of SH has a delayed increase of X-ray flux vs. orbital phase, but the remaining three orbits have an early increase.
This implies the average orbital light-curve over the full SH state is a combination of different shapes. 
This is further verified by the amalgamation of SH orbital light-curves presented by \cite{2011ApJ...736...74L} (their Fig. 6) which show a wide range of count rates during SH. 
The average count rate for SH from RXTE/PCA in that work is consistent with that determined here from BAT, MAXI and ASM data.

\subsection{Main-High State}

The 35-day phase averaged (phase 0.9 to 1.1) orbital light-curve is shown in Figure \ref{fig:Orb_lc_BM123} (panels c and d). 
It is consistent with previously measured MH orbital light-curves \citep{1999ApJ...510..974S,2011ApJ...736...74L}, but here is measured in several energy bands. 
We used the multi-band light-curves, subdivided into sub-states of MH,
to derive the absorption column density, $N_H$, and the transmission fraction, $f$, vs. orbital phase (Figure \ref{fig:Orb_lc_fitMH_B_NH_f}, right column). 

The main results from the current measurement of $N_H$ and $f$ are: 
i) $N_H$ of the time-average orbital light-curve changes systematically with 35-day phase; 
ii) $f$  is consistent with 1 for the orbital phases that are not affected by egress, dip or ingress. 
$N_H$ vs. orbital phase throughout MH is measured here for the first time.
$f$ has been measured previously for short intervals of time using X-ray spectra, and is measured for the first time here for all of MH. 
We find that $N_H$ increases are invariably accompanied by $f$ decreases.
This shows that dips are accompanied by a significant decrease in energy-independent transmission, 
i.e. there is optically thick matter in addition to cold matter absorption associated with the dips.

The $N_H$ and $f$ measurements and the light-curve fits (shown in Figure \ref{fig:Orb_lc_fitMH_B_NH_f}) both show the marching of dips to earlier orbital phase as 35-day phase increases. 
Thus we have measured clearly the marching of dips in fine (0.05) 35-day phase intervals. 
This verifies the previous measurement of orbital phases of individual dips vs. 35-day phase (Fig. 9 of \cite{2011ApJ...736...74L}).
The dip marching can be explained by the accretion stream-disc impact model (\cite{2012MNRAS.425....8I}, see their Fig. 6).
It can also be explained by the dips model of \cite{1999A&A...348..917S}, which involves an out-of-plane stream caused by the asymmetric illumination of HZ Her of X-rays shadowed by the precessing accretion disc.

The orbital phase distribution of dips during MH was measured from RXTE/PCA observations by \citet{2011ApJ...736...74L} (their Fig. 5). 
That work showed that the dips are not uniformly spaced in orbital phase, but cluster mainly after orbital phase 0.7 with a second smaller cluster near phase 0.55 (see their Fig. 7).
This is consistent with what we find (Figure \ref{fig:Orb_lc_fitMH_B_NH_f} both columns),
but the current work shows, for the first time, the evolution of the dip properties with 35-day phase.

Observations with higher time resolution and sensitivity were analyzed by previous studies to see details and variability of individual dips. 
E.g., \citet{2011MNRAS.418.2283I} analyzed RXTE/PCA data and gave three types of pre-eclipse dips with a variety of time durations.
I.e. individual dips are often only minutes in duration, and their occurrence is scattered in orbital phase \citep{2011ApJ...736...74L}, so that the long-term time average dips that we measure is a combination of dip and no-dip data.
Thus our measured width is wider than individual dips and the measured $N_H$ and $f$ are lower and higher, respectively, than that for individual dips.
The advantage of the current analysis is that it gives the systematic (time-average) behaviour of the dips.

Previous studies found the column density during pre-eclipse dip to be on the order of $10^{24} \; \text{cm}^{-2}$ \citep{1995A&A...297..747R, 1999A&A...342..736S, 2005MNRAS.361.1393I}. 
The column density we derived is lower by a factor of ten.
This is not too surprising, because individual dips vary in duration and orbital phase, and averaging of dozens of dips yields a smaller $N_H$. 

When comparing the pre-eclipse dips with the anomalous dips statistically, we find the maximum column density to be about 6.7 times higher in pre-eclipse dips near MH peak. 
During MH decline, the value drops to 3.1, but pre-eclipse dips still have higher maximum column density. 
In comparison, \citet{1995A&A...297..747R} found the factor to be 2.5 times higher column density during pre-eclipse dips. 
We confirm their result of weaker anomalous dips. 
It is reasonable to see a larger factor between the two types of dips from our analysis, as pre-eclipse dips are less scattered and produce higher maximum when averaged. 
Anomalous dips are in general shorter in duration as well \citep{1995A&A...297..747R}. 

In spite of the limitations of our long-term observations, previous references support our conclusion of clustering of absorption dips. 
Individual pieces of observations such as the two anomalous dips observed by EXOSAT in 1984 and 1985 \citep{1995A&A...297..747R} are indeed observed near orbital phase 0.6 within the 35-day phase range of MH-a, among many others. 
A statistical study of RXTE/ASM data \citep{1999ApJ...510..974S} indicates regular absorption dips between orbital phase 0.4 and 0.6 during MH (see their Fig. 5), where several anomalous dips can be seen.  
\citet{2011ApJ...736...74L} find longer in-dip time in this orbital phase range for both MH and SH. 
In particular, their Fig. 9 indicates dip at TO from orbital phase 0.3 to 0.7, which is not seen during most of MH. 
Panel a and b in our Figure \ref{fig:Orb_lc_BM123} produce the same result. 
 
 \subsection{Other remarks}
 
Eclipse egress and ingress are studied at higher time resolution and sensitivity in previous studies (e.g. \cite{1995MNRAS.276..607L,1995ApJ...450..339L,2014ApJ...793...79L,2015ApJ...800...32L}), and those are better suited to measuring eclipse properties.
Here we included egress and ingress in our light-curve model in order to extract the properties of the dips, so there are no new results on egress and ingress in this paper.

Early studies on the 35-day cycle of Her X-1 often found the TO to cluster near orbital phase 0.2 and 0.7 \citep{1999ApJ...510..974S}, which has been rejected by more recent observational studies \citep{2010ApJ...713..318L, 2020ApJ...902..146L}.  
With the average orbital light-curves we present, both can be logically explained. 
The tendency of 35-day TOs near orbital phase 0.7 is possibly due to the regular presence of pre-eclipse dips during early MH and TO, when the observed rate increases after dip. 
In contrast, TOs near orbital phase 0.2 are more likely to be seen when eclipse egress reveals the neutron star to the observer. 

A recent model for dips was proposed by \citet{2012MNRAS.425....8I}. 
The model gives good predictions on the marching of pre-eclipse dips, as well as the width in orbital phase. 
However, the model predicts a dip marching from orbital phase 1.0 to 0.85 as 35-day phase increases from MH-a, which does not align with the observations and occurs late by 0.15 in orbital phase. 
This can be seen by comparing their model 35-day phase vs. orbital plot (their Fig. 6) with
the dips data plot (Fig. 9 of \cite{2011ApJ...736...74L}):
the observed dips are shifted earlier (to the lower left).
We have carried out test calculations that indicate that the disagreement cannot be resolved by simple changes to the disc shape. 
Thus we suggest a different shape of the accretion stream than used by \citet{2012MNRAS.425....8I} (more curved) 
or the out-of-plane stream model of \citet{1999A&A...348..917S}. One of these models is likely better at matching the observed orbital phase vs. 35-day phase diagram for dips.
We plan to carry out such a detailed dips model study in future.

\section{Summary and Conclusion} \label{sec:conclusion}

In this paper, we analyze broadband X-ray observations from the binary system Her X-1/HZ Her with MAXI (2-20 keV; three bands) and Swift/BAT (15-50 keV) data. 
We subdivide the super-orbital 35-day cycle of Her X-1/HZ Her into six states and present long-term average orbital light-curves in multiple energy bands of X-rays. 
We present the most complete set of orbital light-curves of Her X-1 in several X-ray bands, and the change in these light-curves with 35-day phase (Figure \ref{fig:Orb_lc_BM123} here). 
These light-curves will serve as valuable input for modelling the accretion disc structure, which is the cause of the light-curve
changes as the disc rotates with its 35-day period.

The orbital light-curves during LS1 and LS2 are consistent with each other in X-rays. 
We newly report an offset of the optical light-curve later than X-ray by a few hours during LS. 
The orbital light-curves during SH are asymmetric with binary orbit, which is explained by time averaging of rapidly varying light-curves. 
The MH light-curves have the best noise-to-signal, so we further divide them into four sub-states. 

Pre-eclipse dips are studied during the MH sub-states and DEC. 
We confirm many previously known results (e.g. summarized in  \cite{2011ApJ...736...74L}).
Marching of pre-eclipse dips towards earlier 35-day phase is confirmed. 
We find the dips shallower in later 35-day phase and in higher energy bands. 
Anomalous dips are found to cluster near orbital phase 0.6 during early MH. 
We confirm that anomalous dips distinct from and is weaker than the pre-eclipse dips.
The new result is the systematic change in average dip properties with 35-day phase, as summarized in Figure \ref{fig:Orb_lc_fitMH_B_NH_f} above.

\acknowledgments

This research is supported by a grant from the Natural Sciences and Engineering Research Council of Canada. 
This research has made use of the MAXI data provided by RIKEN, JAXA and the MAXI team.


\bibliography{HerX1arXiv}{}
\bibliographystyle{aasjournal}



\newpage
\appendix
\counterwithin{figure}{section}

\section{Orbital Light-curves of RXTE/ASM and MAXI} \label{apdx:Orb_lc}

\begin{figure*}[h]
\gridline{\fig{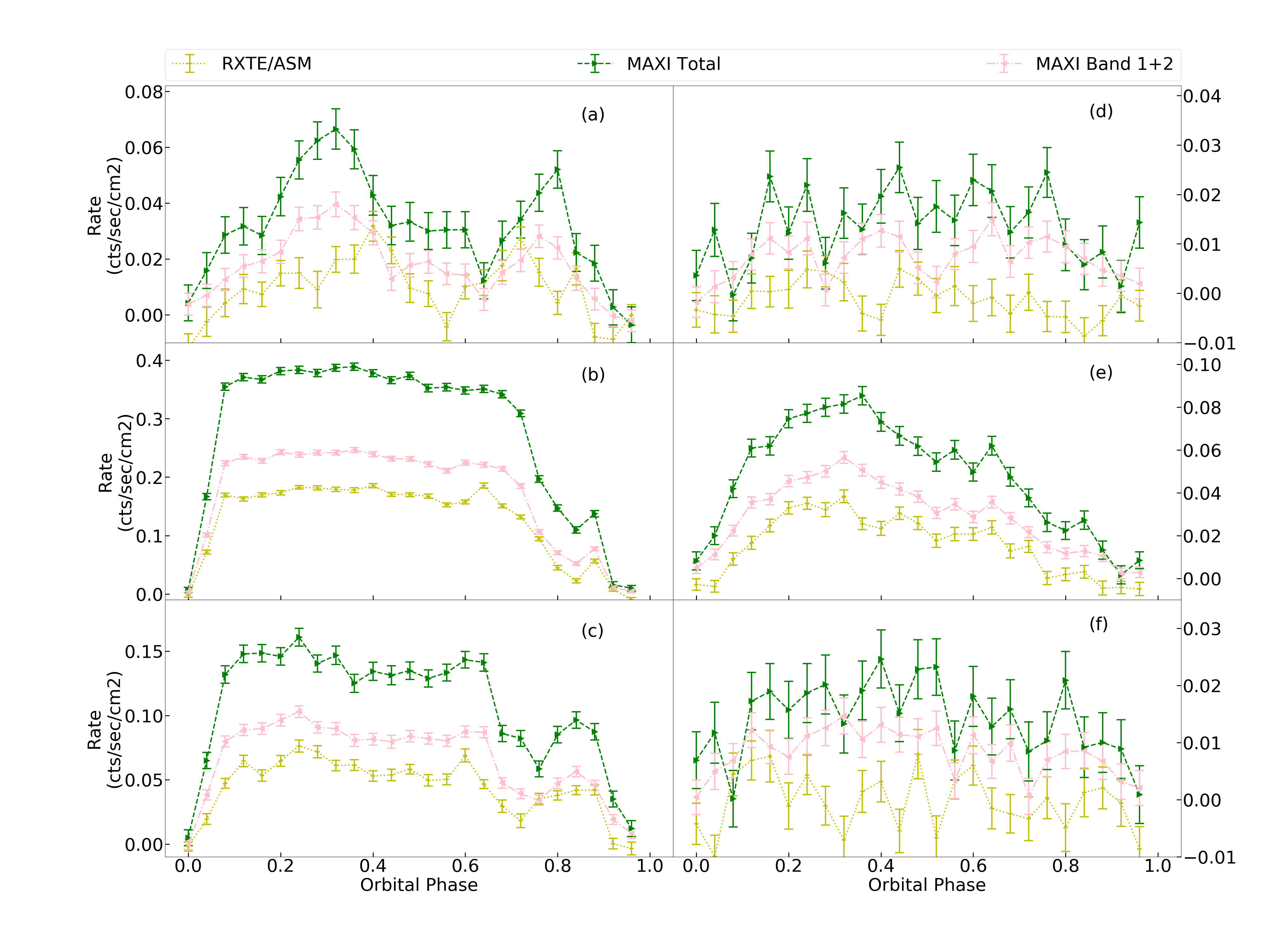}{\textwidth}{}
          }
\caption{Orbital phase light-curves (25 bins) of RXTE/ASM (2-12 keV), MAXI Total (2-20 keV) and Band 1+2 (2-10 keV) during all six 35-day states: 
(a) TO ($ 0.8 \le \phi_{35d} < 0.9$); (b) MH ($ 0.9 \le \phi_{35d} < 1.1$); (c) DEC ($ 0.1 \le \phi_{35d} < 0.22$); (d) LS1 ($ 0.22 \le \phi_{35d} < 0.4$); (e) SH ($ 0.4 \le \phi_{35d} < 0.65$); (f) LS2 ($ 0.65 \le \phi_{35d} < 0.8$).
The left y-axis scale is for the panels in the left column, and the right y-axis scale is for the right panels. }
\label{fig:Orb_lc_AMtM12}
\end{figure*}

Figure \ref{fig:Orb_lc_AMtM12} shows the orbital light-curves derived from the RXTE/ASM data for the six states of the 35-day cycle.
For comparison we show the MAXI 2-20 keV and MAXI 2-10 keV light-curves on the same plot.

\section{Orbital Light-curve Fits during MH and DEC for RXTE/ASM and MAXI} \label{apdx:Orb_lc_fitMH_MA}

\begin{sidewaysfigure}
\includegraphics[angle=0, width=\textwidth]{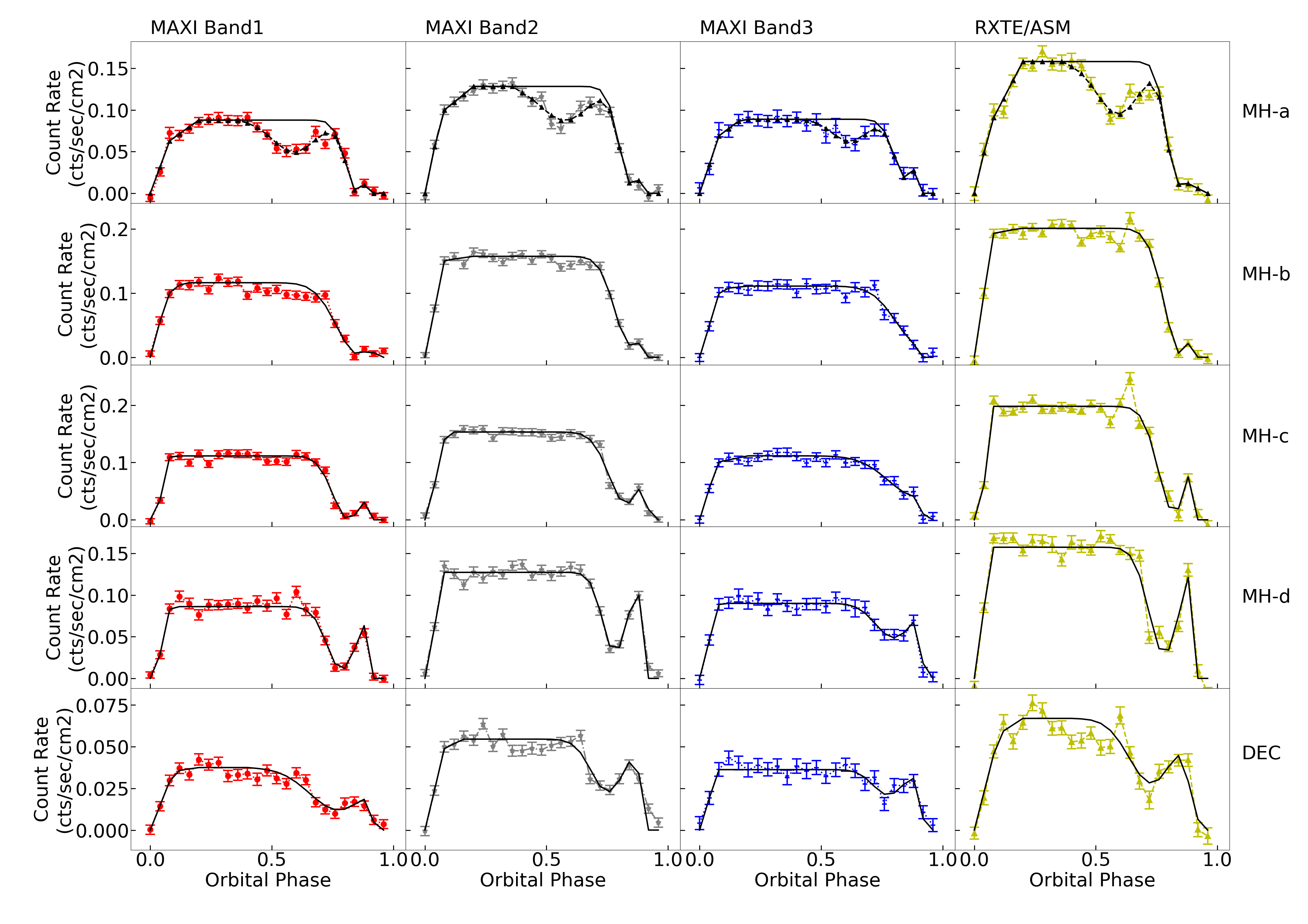}
\caption{Best fit of averaged orbital light-curves during MH sub-states and DEC for MAXI Band 1 (2-4 keV), Band 2  (4-10 keV), Band 3 (10-20 keV) and RXTE/ASM (12-20 keV).  
35-day states: MH-a ($ 0.9 \le \phi_{35d} < 0.95$), MH-b ($ 0.95 \le \phi_{35d} < 1.0$), MH-c ($ 0.0 \le \phi_{35d} < 0.05$), MH-d ($ 0.05 \le \phi_{35d} < 0.1$), and DEC ($ 0.1 \le \phi_{35d} < 0.22$). 
The light-curves are plotted on the same scale for each state.}
\label{fig:Orb_lc_fit_MA}
\end{sidewaysfigure}

The light-curve model fits, described in Section \ref{subsec:fitMH} above,
are shown for the MAXI Band 1, Band 2
and Band 3 light-curves and the RXTE/ASM light-curves in Figure~\ref{fig:Orb_lc_fit_MA}.

\section{Best-fit Parameters of Orbital Light-curves during MH} \label{apdx:fitMH_param}

As the $\chi^2$ values of the best fit functions in Table \ref{tab:chi2} show, introducing a second Gaussian dip greatly improves the fit of orbital light-curves during MH-a. 
The rest of the 35-day sub-states are fit well without a second dip. 

\begin{table}[h]
\centering 
\caption{$\chi^2$ values of the best fit functions during MH Sub-states and DEC. }
\label{tab:chi2}
\begin{tabular}{c c c c c c }
    \hline \hline
     & RXTE & MAXI & MAXI & MAXI & Swift \\
     & /ASM & Band 1 & Band 2 & Band 3 & /BAT \\
    \hline
    MH-a & 37.882 & 25.541 & 30.985 & 11.202 & 133.220 \\
    (2 dips) &&&&&\\
    MH-a & 397.674 & 202.014 & 253.042 & 51.706 & 632.858 \\
    MH-b & 49.609 & 67.078 & 29.729 & 21.430 & 202.647 \\
    MH-c & 66.583 & 32.544 & 35.817 & 20.989 & 302.852 \\
    MH-d & 84.018 & 30.968 & 36.111 & 12.493 & 247.258 \\
    DEC & 90.376 & 32.305 & 71.286 & 17.701 & 76.304 \\
    \hline 
\multicolumn{6}{p{.5\textwidth}}{
\textbf{Note.} 
RXTE/ASM and MAXI have 25-bin orbital light-curves, while Swift/BAT has 50 bins. 
}
\end{tabular}
\end{table}

Table \ref{tab:MHfit_param_A_dip}
gives the best fit parameters and their 1$\sigma$ uncertainties for the light-curve fits for the five sub-states and five energy bands.

\startlongtable
\begin{deluxetable*}{llrlrlrlrlr}
\tabletypesize{\scriptsize}
\tablewidth{0pt}
\tablecaption{The Five Sub-states: Best-fit Parameters (value) and 1$\sigma$ Uncertainties (unc.)\label{tab:MHfit_param_A_dip}}
\tablehead{   \colhead{}      & \colhead{MH-a} & \colhead{} & \colhead{MH-b} & \colhead{} & \colhead{MH-c} & \colhead{} & \colhead{MH-d} & \colhead{} & \colhead{DEC} & \colhead{} \\
 \colhead{parameter} & \colhead{value} & \colhead{unc.} &\colhead{value} & \colhead{unc.} & \colhead{value} & \colhead{unc.}& \colhead{value} & \colhead{unc.} & \colhead{value}& \colhead{unc.}}
\startdata
\cutinhead{$A_{Dip}$}
     \multirow{2}{*}{RXTE/ASM} & \multirow{2}{*}{0.880} &-0.080& \multirow{2}{*}{0.980} &-0.020& \multirow{2}{*}{0.940} &-0.060& \multirow{2}{*}{0.820} &-0.060& \multirow{2}{*}{0.540} &-0.080\\ 
                                                     & &+0.100& &+0.020& &+0.060& &+0.040& &+0.140\\ 
 \multirow{2}{*}{MAXI/Band1} & \multirow{2}{*}{1.000} &-0.060& \multirow{2}{*}{0.960} &-0.060& \multirow{2}{*}{1.000} &-0.060& \multirow{2}{*}{0.880} &-0.060& \multirow{2}{*}{0.680} &-0.060\\ 
                                                     & &+0.000& &+0.040& &+0.000& &+0.080& &+0.100\\ 
 \multirow{2}{*}{MAXI/Band2} & \multirow{2}{*}{0.920} &-0.040& \multirow{2}{*}{0.880} &-0.060& \multirow{2}{*}{0.820} &-0.040& \multirow{2}{*}{0.760} &-0.060& \multirow{2}{*}{0.580} &-0.080\\ 
                                                     & &+0.040& &+0.040& &+0.040& &+0.040& &+0.080\\ 
 \multirow{2}{*}{MAXI/Band3} & \multirow{2}{*}{0.800} &-0.080& \multirow{2}{*}{0.800} &-0.140& \multirow{2}{*}{0.640} &-0.180& \multirow{2}{*}{0.460} &-0.080& \multirow{2}{*}{0.420} &-0.160\\ 
                                                     & &+0.080& &+0.200& &+0.140& &+0.100& &+0.180\\ 
 \multirow{2}{*}{Swift/BAT} & \multirow{2}{*}{0.640} &-0.020& \multirow{2}{*}{0.640} &-0.040& \multirow{2}{*}{0.480} &-0.020& \multirow{2}{*}{0.420} &-0.020& \multirow{2}{*}{0.260} &-0.040\\ 
                                                     & &+0.020& &+0.140& &+0.020& &+0.020& &+0.040\\ 
\cutinhead{$\mu$}
 \multirow{2}{*}{RXTE/ASM} & \multirow{2}{*}{0.850} &-0.000& \multirow{2}{*}{0.851} &-0.006& \multirow{2}{*}{0.822} &-0.000& \multirow{2}{*}{0.781} &-0.000& \multirow{2}{*}{0.723} &-0.017\\ 
                                                     & &+0.003& &+0.006& &+0.006& &+0.006& &+0.023\\ 
 \multirow{2}{*}{MAXI/Band1} & \multirow{2}{*}{0.856} &-0.006& \multirow{2}{*}{0.857} &-0.012& \multirow{2}{*}{0.816} &-0.006& \multirow{2}{*}{0.787} &-0.006& \multirow{2}{*}{0.775} &-0.023\\ 
                                                     & &+0.003& &+0.035& &+0.006& &+0.006& &+0.023\\ 
 \multirow{2}{*}{MAXI/Band2} & \multirow{2}{*}{0.850} &-0.000& \multirow{2}{*}{0.845} &-0.006& \multirow{2}{*}{0.828} &-0.006& \multirow{2}{*}{0.781} &-0.006& \multirow{2}{*}{0.752} &-0.006\\ 
                                                     & &+0.003& &+0.017& &+0.000& &+0.000& &+0.029\\ 
 \multirow{2}{*}{MAXI/Band3} & \multirow{2}{*}{0.850} &-0.000& \multirow{2}{*}{0.903} &-0.046& \multirow{2}{*}{0.897} &-0.075& \multirow{2}{*}{0.799} &-0.017& \multirow{2}{*}{0.775} &-0.029\\ 
                                                     & &+0.008& &+0.064& &+0.035& &+0.029& &+0.035\\ 
 \multirow{2}{*}{Swift/BAT} & \multirow{2}{*}{0.864} &-0.000& \multirow{2}{*}{0.903} &-0.017& \multirow{2}{*}{0.833} &-0.000& \multirow{2}{*}{0.793} &-0.000& \multirow{2}{*}{0.851} &-0.035\\ 
                                                     & &+0.003& &+0.052& &+0.006& &+0.000& &+0.046\\ 
\cutinhead{$\sigma$}
   \multirow{2}{*}{RXTE/ASM} & \multirow{2}{*}{0.050} &-0.001& \multirow{2}{*}{0.068} &-0.006& \multirow{2}{*}{0.064} &-0.004& \multirow{2}{*}{0.062} &-0.004& \multirow{2}{*}{0.106} &-0.034\\ 
                                                     & &+0.000& &+0.004& &+0.006& &+0.006& &+0.014\\ 
 \multirow{2}{*}{MAXI/Band1} & \multirow{2}{*}{0.050} &-0.006& \multirow{2}{*}{0.090} &-0.014& \multirow{2}{*}{0.064} &-0.006& \multirow{2}{*}{0.060} &-0.006& \multirow{2}{*}{0.120} &-0.038\\ 
                                                     & &+0.000& &+0.030& &+0.008& &+0.008& &+0.000\\ 
 \multirow{2}{*}{MAXI/Band2} & \multirow{2}{*}{0.050} &-0.001& \multirow{2}{*}{0.064} &-0.008& \multirow{2}{*}{0.070} &-0.008& \multirow{2}{*}{0.050} &-0.004& \multirow{2}{*}{0.068} &-0.010\\ 
                                                     & &+0.000& &+0.014& &+0.004& &+0.006& &+0.036\\ 
 \multirow{2}{*}{MAXI/Band3} & \multirow{2}{*}{0.050} &-0.005& \multirow{2}{*}{0.098} &-0.032& \multirow{2}{*}{0.120} &-0.056& \multirow{2}{*}{0.074} &-0.018& \multirow{2}{*}{0.062} &-0.032\\ 
                                                     & &+0.000& &+0.022& &+0.000& &+0.036& &+0.050\\ 
 \multirow{2}{*}{Swift/BAT} & \multirow{2}{*}{0.050} &-0.002& \multirow{2}{*}{0.092} &-0.012& \multirow{2}{*}{0.064} &-0.004& \multirow{2}{*}{0.052} &-0.002& \multirow{2}{*}{0.104} &-0.026\\ 
                                                     & &+0.000& &+0.026& &+0.004& &+0.002& &+0.016\\ 
\cutinhead{$\phi_0$}
 \multirow{2}{*}{RXTE/ASM} & \multirow{2}{*}{0.930} &-0.000& \multirow{2}{*}{0.910} &-0.000& \multirow{2}{*}{0.910} &-0.000& \multirow{2}{*}{0.910} &-0.000& \multirow{2}{*}{0.930} &-0.000\\ 
                                                     & &+0.000& &+0.000& &+0.000& &+0.000& &+0.020\\ 
 \multirow{2}{*}{MAXI/Band1} & \multirow{2}{*}{0.910} &-0.000& \multirow{2}{*}{0.930} &-0.020& \multirow{2}{*}{0.910} &-0.000& \multirow{2}{*}{0.910} &-0.000& \multirow{2}{*}{0.930} &-0.020\\ 
                                                     & &+0.000& &+0.020& &+0.000& &+0.000& &+0.000\\ 
 \multirow{2}{*}{MAXI/Band2} & \multirow{2}{*}{0.910} &-0.000& \multirow{2}{*}{0.910} &-0.000& \multirow{2}{*}{0.930} &-0.000& \multirow{2}{*}{0.910} &-0.000& \multirow{2}{*}{0.910} &-0.000\\ 
                                                     & &+0.000& &+0.020& &+0.000& &+0.000& &+0.020\\ 
 \multirow{2}{*}{MAXI/Band3} & \multirow{2}{*}{0.910} &-0.000& \multirow{2}{*}{0.910} &-0.000& \multirow{2}{*}{0.930} &-0.020& \multirow{2}{*}{0.930} &-0.000& \multirow{2}{*}{0.930} &-0.000\\ 
                                                     & &+0.000& &+0.040& &+0.000& &+0.000& &+0.020\\ 
 \multirow{2}{*}{Swift/BAT} & \multirow{2}{*}{0.910} &-0.000& \multirow{2}{*}{0.910} &-0.000& \multirow{2}{*}{0.910} &-0.000& \multirow{2}{*}{0.910} &-0.000& \multirow{2}{*}{0.910} &-0.000\\ 
                                                     & &+0.000& &+0.000& &+0.000& &+0.000& &+0.000\\ 
\cutinhead{$\phi_2$}
     \multirow{2}{*}{RXTE/ASM} & \multirow{2}{*}{0.070} &-0.000& \multirow{2}{*}{0.078} &-0.008& \multirow{2}{*}{0.078} &-0.008& \multirow{2}{*}{0.070} &-0.000& \multirow{2}{*}{0.101} &-0.031\\ 
                                                     & &+0.031& &+0.010& &+0.000& &+0.013& &+0.034\\ 
 \multirow{2}{*}{MAXI/Band1} & \multirow{2}{*}{0.080} &-0.010& \multirow{2}{*}{0.070} &-0.000& \multirow{2}{*}{0.070} &-0.000& \multirow{2}{*}{0.078} &-0.008& \multirow{2}{*}{0.096} &-0.026\\ 
                                                     & &+0.034& &+0.034& &+0.008& &+0.021& &+0.036\\ 
 \multirow{2}{*}{MAXI/Band2} & \multirow{2}{*}{0.070} &-0.000& \multirow{2}{*}{0.080} &-0.010& \multirow{2}{*}{0.075} &-0.005& \multirow{2}{*}{0.075} &-0.005& \multirow{2}{*}{0.080} &-0.010\\ 
                                                     & &+0.023& &+0.010& &+0.023& &+0.010& &+0.021\\ 
 \multirow{2}{*}{MAXI/Band3} & \multirow{2}{*}{0.080} &-0.010& \multirow{2}{*}{0.086} &-0.016& \multirow{2}{*}{0.073} &-0.003& \multirow{2}{*}{0.075} &-0.005& \multirow{2}{*}{0.073} &-0.003\\ 
                                                     & &+0.060& &+0.021& &+0.031& &+0.021& &+0.034\\ 
 \multirow{2}{*}{Swift/BAT} & \multirow{2}{*}{0.070} &-0.000& \multirow{2}{*}{0.070} &-0.000& \multirow{2}{*}{0.070} &-0.000& \multirow{2}{*}{0.070} &-0.000& \multirow{2}{*}{0.070} &-0.000\\ 
                                                     & &+0.003& &+0.000& &+0.000& &+0.000& &+0.008\\ 
\cutinhead{$\phi_3$}
 \multirow{2}{*}{RXTE/ASM} & \multirow{2}{*}{0.200} &-0.024& \multirow{2}{*}{0.188} &-0.118& \multirow{2}{*}{0.078} &-0.008& \multirow{2}{*}{0.078} &-0.008& \multirow{2}{*}{0.200} &-0.120\\ 
                                                     & &+0.000& &+0.012& &+0.003& &+0.122& &+0.000\\ 
 \multirow{2}{*}{MAXI/Band1} & \multirow{2}{*}{0.200} &-0.066& \multirow{2}{*}{0.132} &-0.062& \multirow{2}{*}{0.080} &-0.010& \multirow{2}{*}{0.080} &-0.010& \multirow{2}{*}{0.200} &-0.130\\ 
                                                     & &+0.000& &+0.068& &+0.005& &+0.120& &+0.000\\ 
 \multirow{2}{*}{MAXI/Band2} & \multirow{2}{*}{0.200} &-0.055& \multirow{2}{*}{0.200} &-0.130& \multirow{2}{*}{0.083} &-0.013& \multirow{2}{*}{0.078} &-0.008& \multirow{2}{*}{0.159} &-0.089\\ 
                                                     & &+0.000& &+0.000& &+0.117& &+0.122& &+0.041\\ 
 \multirow{2}{*}{MAXI/Band3} & \multirow{2}{*}{0.171} &-0.101& \multirow{2}{*}{0.200} &-0.130& \multirow{2}{*}{0.200} &-0.130& \multirow{2}{*}{0.083} &-0.013& \multirow{2}{*}{0.078} &-0.008\\ 
                                                     & &+0.029& &+0.000& &+0.000& &+0.117& &+0.122\\ 
 \multirow{2}{*}{Swift/BAT} & \multirow{2}{*}{0.200} &-0.062& \multirow{2}{*}{0.200} &-0.052& \multirow{2}{*}{0.200} &-0.130& \multirow{2}{*}{0.200} &-0.130& \multirow{2}{*}{0.127} &-0.057\\ 
                                                     & &+0.000& &+0.000& &+0.000& &+0.000& &+0.073\\ 
\cutinhead{$R_{ratio}$}
 \multirow{2}{*}{RXTE/ASM} & \multirow{2}{*}{0.540} &-0.080& \multirow{2}{*}{0.960} &-0.240& \multirow{2}{*}{0.600} &-0.200& \multirow{2}{*}{0.940} &-0.160& \multirow{2}{*}{0.860} &-0.580\\ 
                                                     & &+0.160& &+0.040& &+0.160& &+0.060& &+0.140\\ 
 \multirow{2}{*}{MAXI/Band1} & \multirow{2}{*}{0.720} &-0.160& \multirow{2}{*}{0.840} &-0.200& \multirow{2}{*}{0.540} &-0.220& \multirow{2}{*}{0.640} &-0.320& \multirow{2}{*}{0.940} &-0.600\\ 
                                                     & &+0.220& &+0.160& &+0.280& &+0.360& &+0.060\\ 
 \multirow{2}{*}{MAXI/Band2} & \multirow{2}{*}{0.760} &-0.080& \multirow{2}{*}{0.960} &-0.260& \multirow{2}{*}{0.760} &-0.220& \multirow{2}{*}{0.920} &-0.240& \multirow{2}{*}{0.900} &-0.360\\ 
                                                     & &+0.140& &+0.040& &+0.240& &+0.080& &+0.100\\ 
 \multirow{2}{*}{MAXI/Band3} & \multirow{2}{*}{0.780} &-0.500& \multirow{2}{*}{0.960} &-0.480& \multirow{2}{*}{0.900} &-0.340& \multirow{2}{*}{0.960} &-0.380& \multirow{2}{*}{0.960} &-0.540\\ 
                                                     & &+0.220& &+0.040& &+0.100& &+0.040& &+0.040\\ 
 \multirow{2}{*}{Swift/BAT} & \multirow{2}{*}{0.900} &-0.100& \multirow{2}{*}{0.940} &-0.020& \multirow{2}{*}{1.000} &-0.000& \multirow{2}{*}{1.000} &-0.060& \multirow{2}{*}{0.900} &-0.120\\ 
                                                     & &+0.040& &+0.020& &+0.000& &+0.000& &+0.100\\ 
\hline
\enddata
\end{deluxetable*}

\end{document}